\newcommand{\ud}{\rm d}
\newcommand{\un}{~\mathrm}

\newcommand{\gravity}{\left(\overrightarrow{e_X},\overrightarrow{e_Z}\right)}
\newcommand{\local}{\left(\overrightarrow{e_x},\overrightarrow{e_z}\right)}

\documentstyle[aps,psfig,epsfig]{revtex}

\begin{document}
\draft
\author{D. Bonamy, F. Daviaud and L. Laurent}
\address{}
\date{\today}
\title{Experimental study of granular surface flows via a fast camera: a continuous description}
\maketitle
\begin{abstract}
Depth averaged conservation equations are written for granular surface flows. Their application to the study of steady surface flows in a rotating drum allows to find experimentally the constitutive relations needed to close these equations from measurements of the velocity profile in the flowing layer at the center of the drum and from the flowing layer thickness and the static/flowing boundary profiles. The velocity varies linearly with depth, with a gradient independent of both the flowing layer thickness and the static/flowing boundary local slope. The first two closure relations relating the flow rate and the momentum flux to the flowing layer thickness and the slope are then deduced. Measurements of the profile of the flowing layer thickness and the static/flowing boundary in the whole drum explicitly give the last relation concerning the force acting on the flowing layer. Finally, these closure relations are compared to existing continuous models of surface flows.
\end{abstract}


\section{introduction}

Granular media present numbers of interesting and unusual properties~\cite{jaeger96}: the intrinsic dissipative nature of the interactions between the constituent macroscopic particles sets granular matter apart from conventional gases, liquids or solids. One of the most interesting phenomenon in granular systems is the transition from a static equilibrium to a granular flow: contrary to ordinary fluids, they can remain static even with an inclined free surface. But when the angle of the surface exceeds some threshold value $\theta_d$, the pile can not sustain the steep surface and starts to flow until its angle relaxes under a given angle of repose $\theta_r$. The motion has the particularity to be a surface flow: most of the pile remains static. The condition governing the transition between the solid state and the liquid state as well as the internal equations of the flowing layer are still today under debate~\cite{mills99}. This makes the description of an avalanche rather difficult.

People have benefited from the particularity of avalanches to be surface flows to apply to them depth averaged conservation laws, the Saint Venant equations~\cite{venant50}, and to use the approximations developed in hydrodynamics for a thin fluid film flowing down an inclined plane. Savage and Hutter~\cite{savage89}, and more recently Pouliquen~\cite{pouliquen99bis} applied these equations to describe the motion of granular material down a rough inclined plane. Khakhar {\it et al.}~\cite{khakhar97}, and T. Elperin and A. Vikhansky~\cite{elperin98} have studied steady flows in a rotating drum using the same theoretical framework. They succeed to report on the different free-surface shapes successively observed when the rotation speed increases. Finally, Douady {\em et al.}~\cite{douady99} and very recently Khakhar {\em et al.}~\cite{khakhar01} have used these equations to describe avalanches at the surface of a heap. Another model, referred to as the BCRE model~\cite{bouchaud94}, has also been proposed a few years ago: it deals with thin surface granular flows including avalanches and succeeds for example to report on the existence of the two angles $\theta_d$ and $\theta_r$. This model has then been extended to describe thick surface flows~\cite{boutreux98,aradian98}.

All these models are based on the assumption that a strict separation between flowing grains and static grains can be made. They end in a system of two coupled equations with the same form, differing only through their closure relations. Those are chosen from theoretical considerations, linking for example the coupling term to the effective force applying on the flowing phase (for Saint-Venant approach) or to the microscopic effects of the flowing grains collisions on the static bed (for BCRE approach). An original approach, proposed by Aranson and Tsimring, should also be mentioned~\cite{aranson00}. Assuming a continuous transition between flowing grains and static grains,  they have developped a theory based on hydrodynamic equations coupled with an order parameter equation describing the solid/flowing transition. This theory accounts for most of the behaviours recently observed in experiments on rough inclined plane~\cite{daerr99}.  

Thin rotating drums, {\em i.e.} rotating drums whose gap is small compared to the diameter,  have been widely used to investigate properties of granular materials: Evesque and Rajchenbach~\cite{evesque88} and Jaeger {\it et al.}~\cite{nagel89} have studied avalanche size distribution. Rajchenbach has used it in an attempt to find constitutive mechanical laws in 2D beads packing~\cite{rajchenbach00}. 
Nakagawa {\it et al.}~\cite{nakagawa93} have used magnetic resonance imaging to measure density and velocity field in a thick rotating drum.
We propose here to study steady surface flows in a rotating drum for a quasi-2D packing of monodisperse beads in order to find experimentally the closure relations needed to complete the depth averaged hydrodynamic equations. 

In section~\ref{S2}, basic conservation laws are presented. It is shown how the closure relations can be deduced from the experimental velocity profile inside the flowing layer at the center of the drum and from the flowing layer profile. The experimental set-up is then described (section~\ref{S3}). Beads are studied on the microscopic scale in section~\ref{S4}: velocity and volume fraction profiles are measured on elementary slices via a fast camera. In section~\ref{S5}, the flowing layer thickness and the angle between the interface and the horizontal are measured in the whole drum. The closure relations are deduced. Finally, they are compared in section~\ref{S6} to the ones assumed by the different models proposed in the litterature. 

\section{theoretical frame: Saint-Venant description} \label{S2}

\begin{figure}
\begin{center}
\psfig{figure=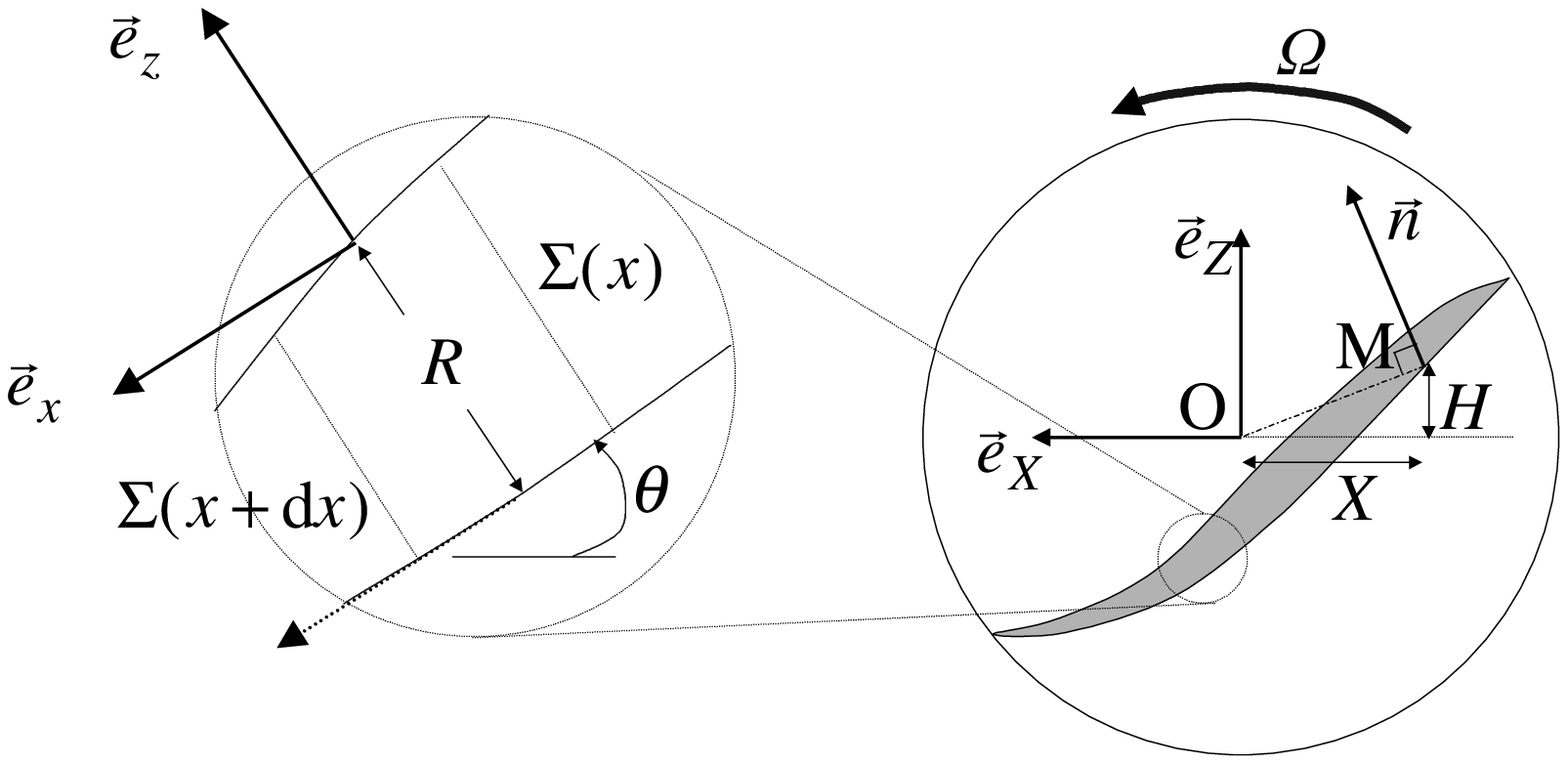, width=0.8\textwidth}
\vspace{0.5cm}
\caption{Bonamy, Phys. Fluids}
\label{f.1}
\end{center}
\end{figure}

Let us consider the situation depicted fig.~\ref{f.1}. The approach is limited to the 2D situation. The material can then be divided into two parts: a thin cascading layer flows on a static bed. The boundary between these two phases is supposed to be sharp. This assumption will be discussed and justified in section~\ref{S4}. As the flowing layer is assumed to be thin, the velocity of grains is essentially parallel to the static/flowing boundary. The most natural frame is consequently the curvilinear frame $\local$ where $\overrightarrow{e_x}$ (resp. $\overrightarrow{e_z}$) is locally tangent (resp. normal) to this interface. The $z$ origin is set at the free surface. Calling $(v_x,v_z)$ the velocity at the $(x,z)$ coordinate, the mass and momentum conservation equations can be written~\cite{douady99,khakhar01}:

\begin{equation}
\frac{\partial}{\partial t} (<\rho> R)+\frac{\partial}{\partial x}  (R<\rho v_x>) +(\rho v_z)|_{z=-R}=0
\label{e.1}
\end{equation}
 
\begin{equation}
\frac{\partial}{\partial t}(R<\rho v_x>)+\frac{\partial}{\partial x} (R<\rho {v_x}^2>)+(\rho v_x v_z)|_{z=-R}=F
\label{e.2}
\end{equation}

\noindent where $R$ is the flowing layer thickness, $\rho$ the solid fraction and $F$ the $x$-component of the force acting on the volume of the flowing layer bounded by the two surfaces $\Sigma(x)$ and $\Sigma(x+\ud x)$. The quantities $(R<\rho v_x>)$ and $(R<\rho {v_x}^2>)$ are respectively the flow rate and the flux of x-component of momentum through the surface $\Sigma(x)$ (see fig.~\ref{f.1}) while $(\rho v_z)|_{z=-R}$ and $(\rho v_x v_z)|_{z=-R}$ represent respectively the flow rate and the flux of x-component of momentum passing through the static/flowing boundary. $<A>$ denotes the average of the quantity $A$ across the flowing layer~:

\begin{equation}
<A>=\frac{1}{R} \int_{-R}^{0} A {\rm d} z
\label{e.3}
\end{equation}

In the following, the volume fraction $\rho$ is supposed to be constant (this assumption will be discussed and justified in section~\ref{S4}) and eq.~(\ref{e.1},\ref{e.2}) are simplified accordingly. The expressions of $<v_x>$, $<{v_x}^2>$ and $F$ as a fonction of the flow macroscopic parameters are required to close the two equations~(\ref{e.1},\ref{e.2}). Translating the packing upwards should not change its dynamical equations. Consequently, the three closure relations can be written as $<v_x>(R,\theta)$, $<{v_x}^2>(R,\theta)$ and $F(R,\theta)$ where $\theta(x,t)$ is the angle made by the solid/flowing interface with the horizontal. Let us apply these two conservation equations to the description of steady surface flows in a rotating drum  (see fig.~\ref{f.1}). In this case, time derivatives terms in eq.~(\ref{e.1},\ref{e.2})  vanish and the velocity $\vec{v}|_{z=-R}$ of the static grains at the bed/layer interface can be written:

\begin{equation}
\vec{v}|_{z=-R}=\Omega r \vec{n}
\label{e.5}
\end{equation}   

\noindent where $\Omega$ is the rotation speed of the drum, $r$ the distance between the center of the drum $\mathrm{O}$ and the point $\mathrm{M}$ of the interface where $\vec{v}|_{z=-R}$ is calculated, and $\vec{n}$ the unity vector perpendicular to $\overrightarrow{\mathrm{OM}}$. Then, eq.~(\ref{e.1},\ref{e.2}) become:

\begin{equation}
\frac{\ud}{\ud x}  (R <v_{x}>)= - \Omega \Big(X\cos\theta+H\sin\theta\Big)
\label{e.6}
\end{equation}

\begin{equation}
\frac{\ud }{\ud x}(R<{v_{x}}^2>)+\Omega^2 \Big(X\cos\theta+H\sin\theta\Big) \Big(X\sin\theta-H\cos\theta\Big).{\mathcal{H}}\Big(X\cos\theta+H\sin\theta\Big) =\frac{F}{\rho}
\label{e.7}
\end{equation}   

\noindent where $(X(x),H(x))$ is the coordinate of $\mathrm{M}$ in the gravity frame whose origin coincides with the center of the drum (see fig.~\ref{f.1}) and $\mathcal{H}$ the Heaviside function.

The aim of the paper is first to investigate the beads behaviour on the microscopic scale: the profiles of the velocity $v_{x}(z)$ and volume fraction $\rho(z)$ are measured on an elementary slice. The evolution of these profiles with $R$ and $\theta$ allows one to deduce the first two closure relations $<{v_{x}}>(R,\theta)$ and $<{v_{x}}^2>(R,\theta)$. Then, profiles $R(x)$ and $\theta(x)$ are measured on the macroscopic scale, in the whole drum. These profiles confirm the form of the first two closure relations, extend their range of validity and allow one to find the last closure relation $F(R,\theta)$.

\section{Experimental setup} \label{S3}

\begin{figure}
\begin{center}
\psfig{figure=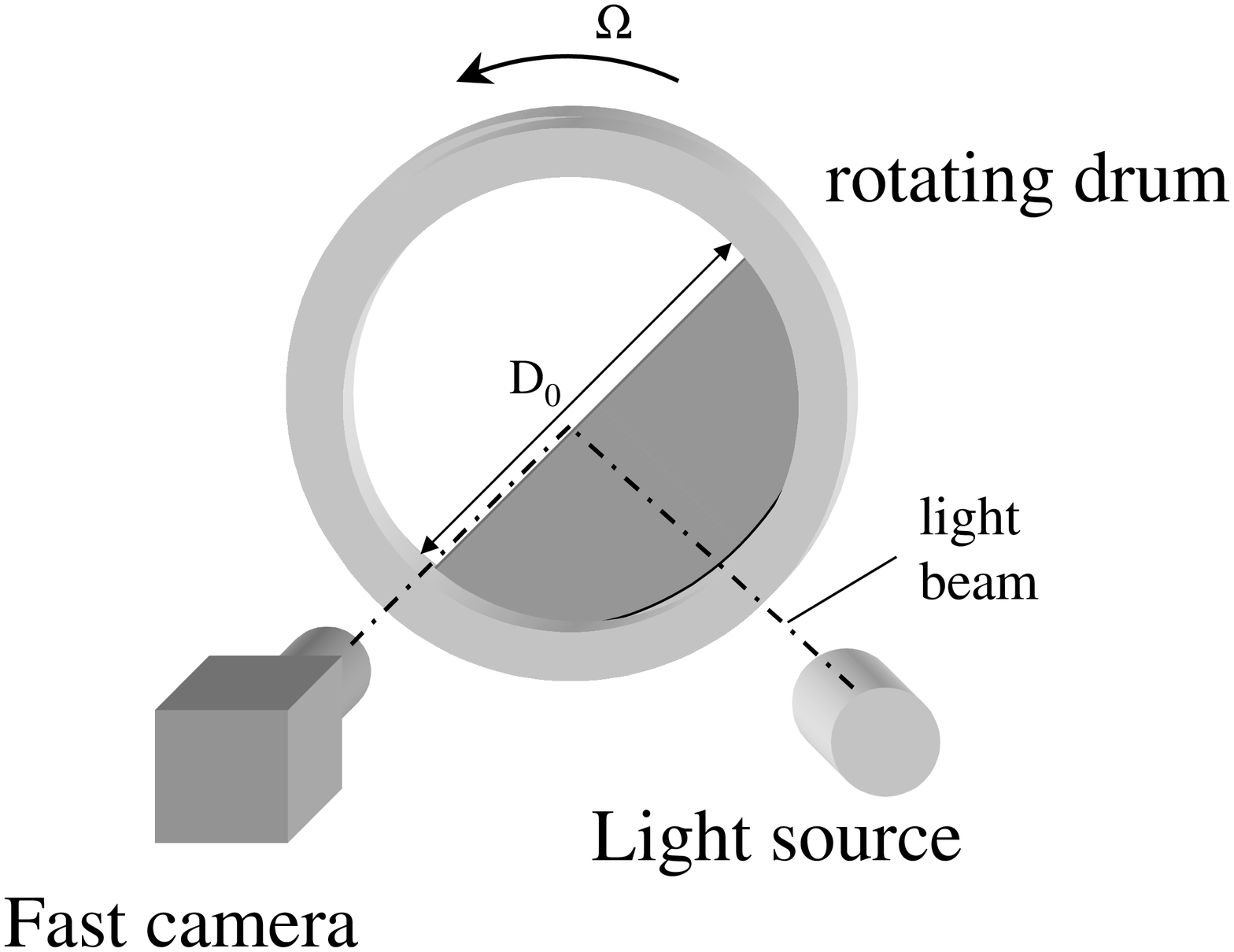, width=0.8\textwidth}
\caption{Bonamy, Phys. Fluids}
\label{f.2}
\end{center}
\end{figure}

The experimental set-up is illustrated in fig.~\ref{f.2}. It consists mainly in a duralumin rotating drum of diameter $D_0=45\un{cm}$ and of gap $e=7\un{mm}$. The drum is half filled by a packing of steel beads of diameter $d=3\pm 0.05\un{mm}$. The drum can then be considered as a Hele Shaw cell ($D_0 \gg e \approx 2d$) where the packing has a 3D microscopic structure (especially a disordered structure contrary to 2D packing whose stable structure is crystalline) while keeping a 2D macroscopic geometry. The volume fraction measured just after the filling is $0.57 \pm 0.02$. The gap $e$ has been chosen slightly larger than $2d$ to prevent any jamming in beads flow. The use of steel beads of millimetric size allows to have a good control of geometrical and mechanical properties and to limit capillary and electrostatic effects at the expense of a larger scale experiment.

The rotation speed $\Omega$ can be varied from $0.001\un{r.p.m.}$ up to $30\un{r.p.m}$. For small $\Omega$ -~typically smaller than $0.1\un{r.p.m.}$~-, the mean angle of the free surface $\bar{\theta}$ oscillates between the angle of flow $\theta_d=27.5 \pm 0.7^\circ$ and the angle of repose $\theta_r=23.5 \pm 0.6^\circ$. This intermittent behaviour is not studied in this paper. For $\Omega \geq 0.1\un{r.p.m.}$, surface flow becomes steady. The beads are lighted via a continuous halogen lamp and regimes obtained for $\Omega$ varying from $1\un{r.p.m.}$ up to $6\un{r.p.m.}$ are recorded via a fast camera. On this range of rotation speed, the width of the flowing layer $R$ is well defined and inertial effects are negligible (the Froude number $Fr=\Omega^2 D_0/2 g \leq 0.01$ where $g$ is the gravity constant).  

\section{microscopic scale: velocity and volume fraction profile on an elementary slice} \label{S4}

\begin{figure}
\centering
\epsfig{figure=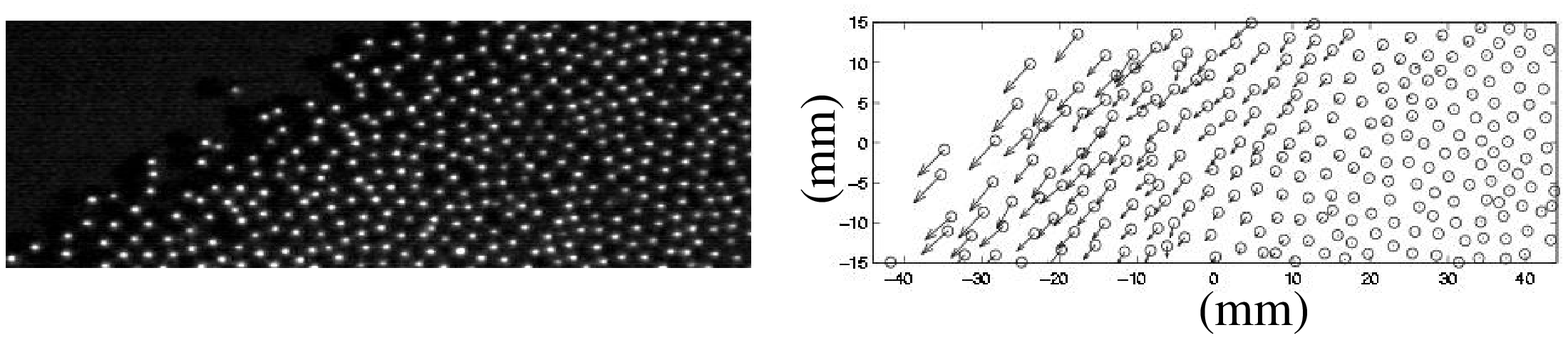, width=\textwidth}
\vspace{0.5cm}
\caption{Bonamy, Phys. Fluids}
\label{f.3}
\end{figure}

Sequences of frames are recorded via a fast camera at a sampling rate $f_s = 1\un{kHz}$. The shutter speed is of $0.1\un{ms}$. The recorded region is located at the point where $(\overrightarrow{v_z}|_{z=-R}.\overrightarrow{n})=0$ (roughly corresponding to the center of the drum). In this region, $\theta$, $R$ and $\{v_{x}(z)\}$ are invariant under a small translation along $\overrightarrow{e}_x$. The recorded region can then be considered as a statisfactory elementary slice. $R$ and $\theta$ in this area are controlled by changing $\Omega$, but can not been modified independently. The recorded region size in pixel is  $480\times234$, one pixel corresponding to $0.227\un{mm}$. The digital images are  processed to obtain the position of the center of mass of the beads seen through the glass porthole (see fig.~\ref{f.3}). The image of a single bead is made up of about $23\pm3$ pixels depending on the distance of the bead to the porthole. The error on the determination of beads location is thus about $50~\mu\mathrm{m}$. One can then extract both volume fraction and velocity profiles.

\begin{figure}
\centering
\begin{minipage}[c]{0.45\textwidth}
\centering\epsfig{figure=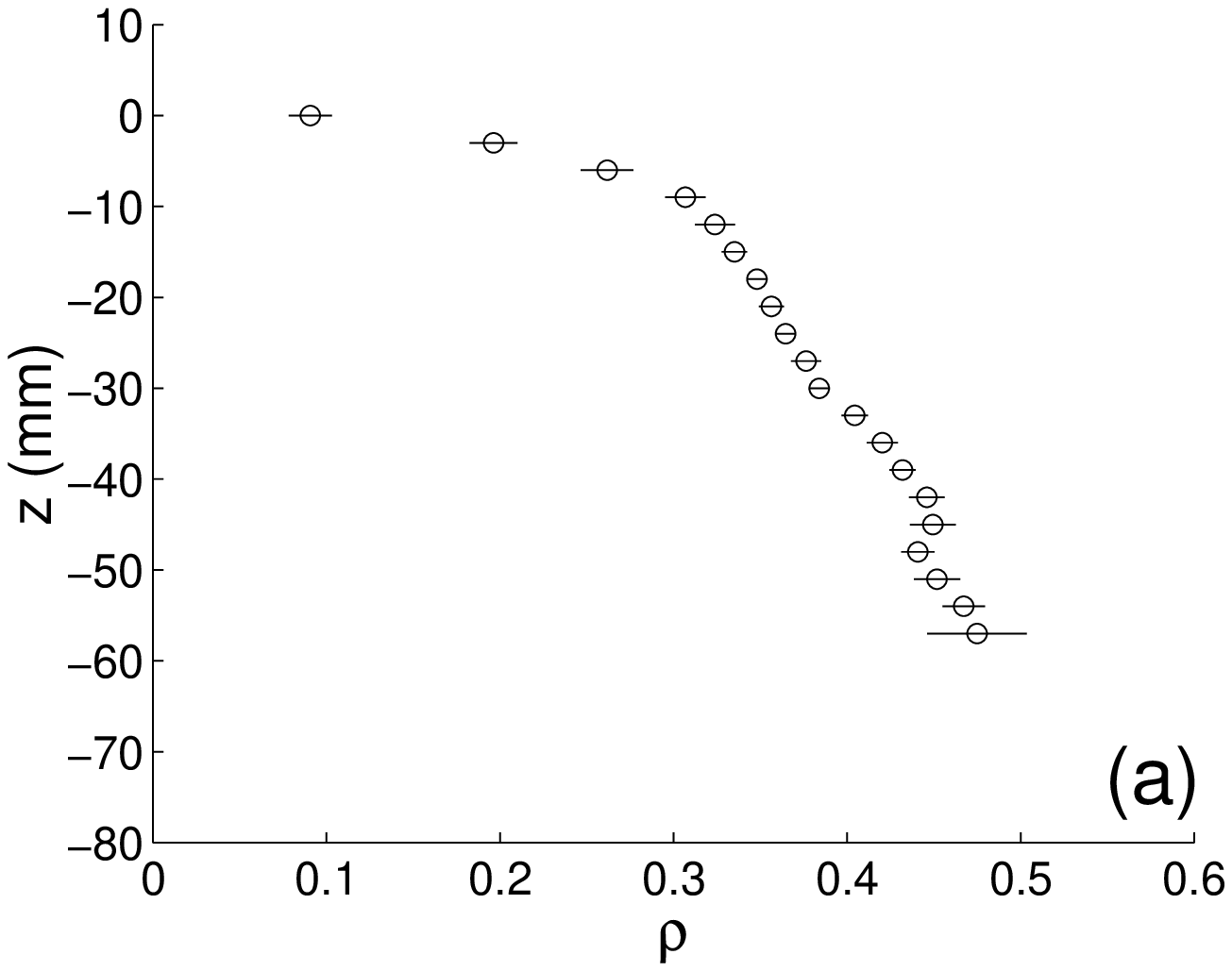, width=\textwidth}
\end{minipage}
\begin{minipage}[c]{0.45\textwidth}
\centering\epsfig{figure=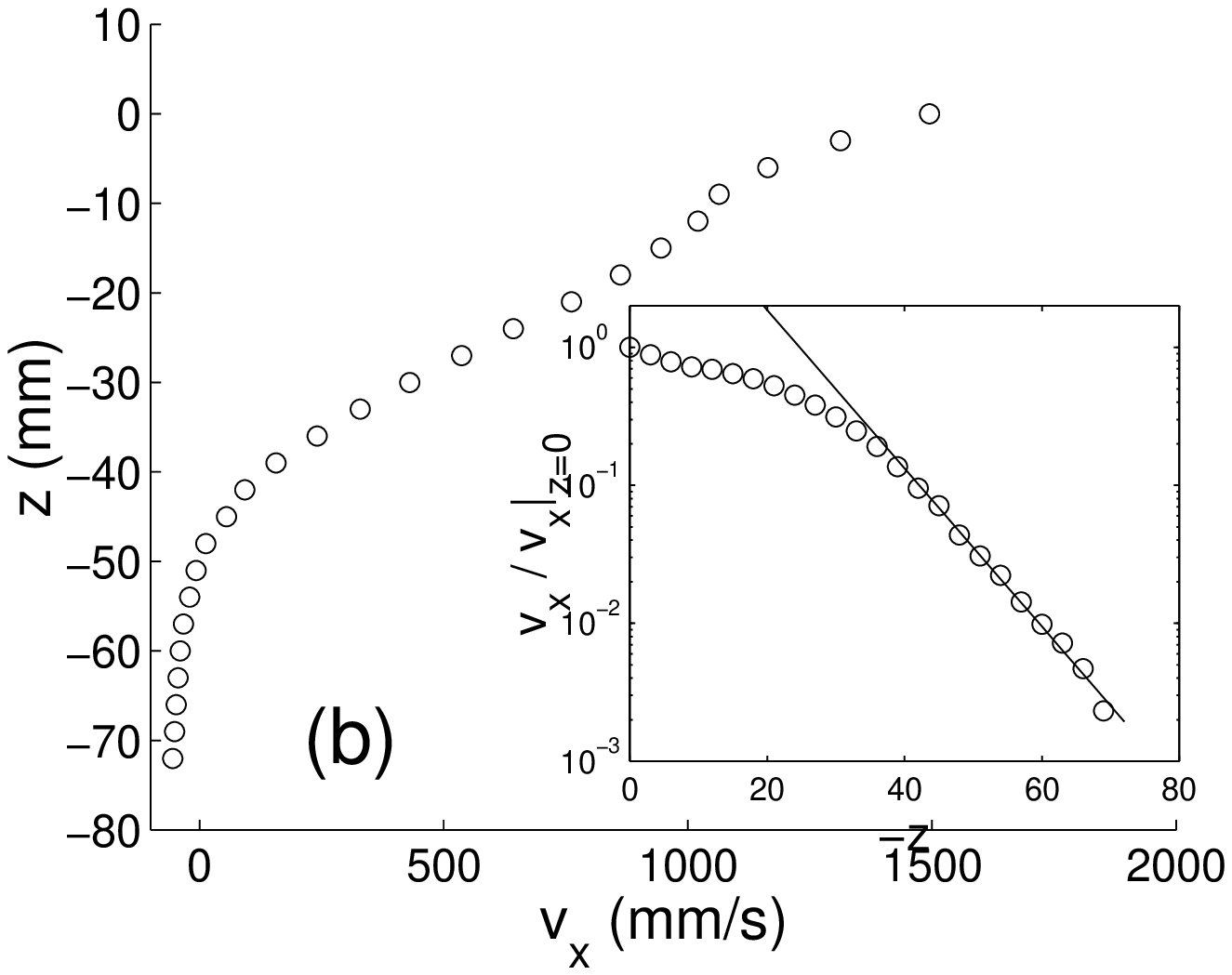, width=\textwidth}
\end{minipage}
\vspace{0.5cm}
\caption{Bonamy, Phys. Fluids}
\label{f.4}
\end{figure}

Volume fraction profile can not be measured directly in our experiment since all beads are not seen through the porthole. One has thus calculated from numerical simulated packing the relationship between the real volume fraction and the volume fraction estimated from the beads seen through the porthole (see appendix A). The experimental volume fraction profiles can then be deduced. The errors made on these profiles is mainly dominated by the statistical processing of the data: profiles have been averaged over $1000$ frames to limit this error.
A typical volume fraction profile is represented fig.~\ref{f.4}a: $\rho$ varies typically from $0.3$ to $0.4$ in the flowing layer, except at the free surface in a small region (about three beads diameters) where $\rho$ drops quickly down to zero. This last region is  assumed to be sharp and corresponds to the free surface boundary. The assumption $\rho \simeq constant$ can then be questioned: relaxing this assumption amounts to replace $<v_{x}>$ and $<{v_{x}}^2>$ respectively by $<\rho v_{x}>/<\rho>$ and $<\rho(z) {v_{x}}^2>/<\rho>$. As $\rho$ variations are negligible compared to $v_{x}$ variations, $\rho$ can be assumed to be constant. 

Frames processing allows to evaluate beads velocity (see appendix B). A typical velocity profile is represented fig.~\ref{f.4}b. The velocity profile is linear in the whole flowing layer. The inset, representing the variation of the velocity vs. depth in semilogarithmic scale, emphasizes the creep motion of the static phase. In this phase, $v_{x}$ decreases exponentially with the depth with a characteristic decay length $\lambda_{creep}=7.6\un{mm} \simeq 2.5d$ in agreement with~\cite{komatsu01}. The creep motion is thus localised in a narrow layer at the solid/liquid interface whose thickness can be considered as zero at macroscopic scale. The creep motion is not studied in this paper.

\begin{figure}
\centering
\begin{minipage}[c]{0.45\textwidth}
\centering\epsfig{figure=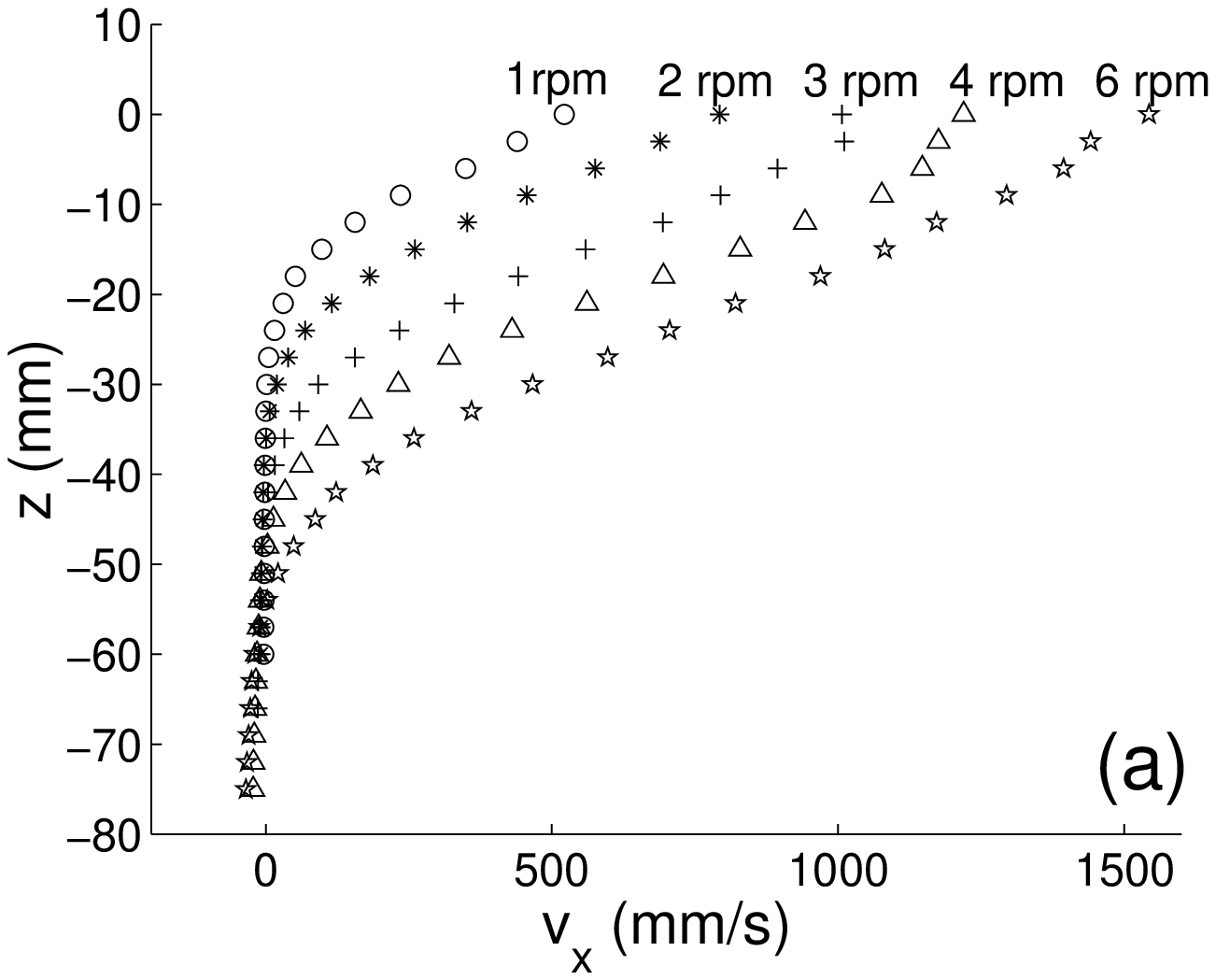, width=\textwidth}
\end{minipage}
\begin{minipage}[c]{0.45\textwidth}
\centering\epsfig{figure=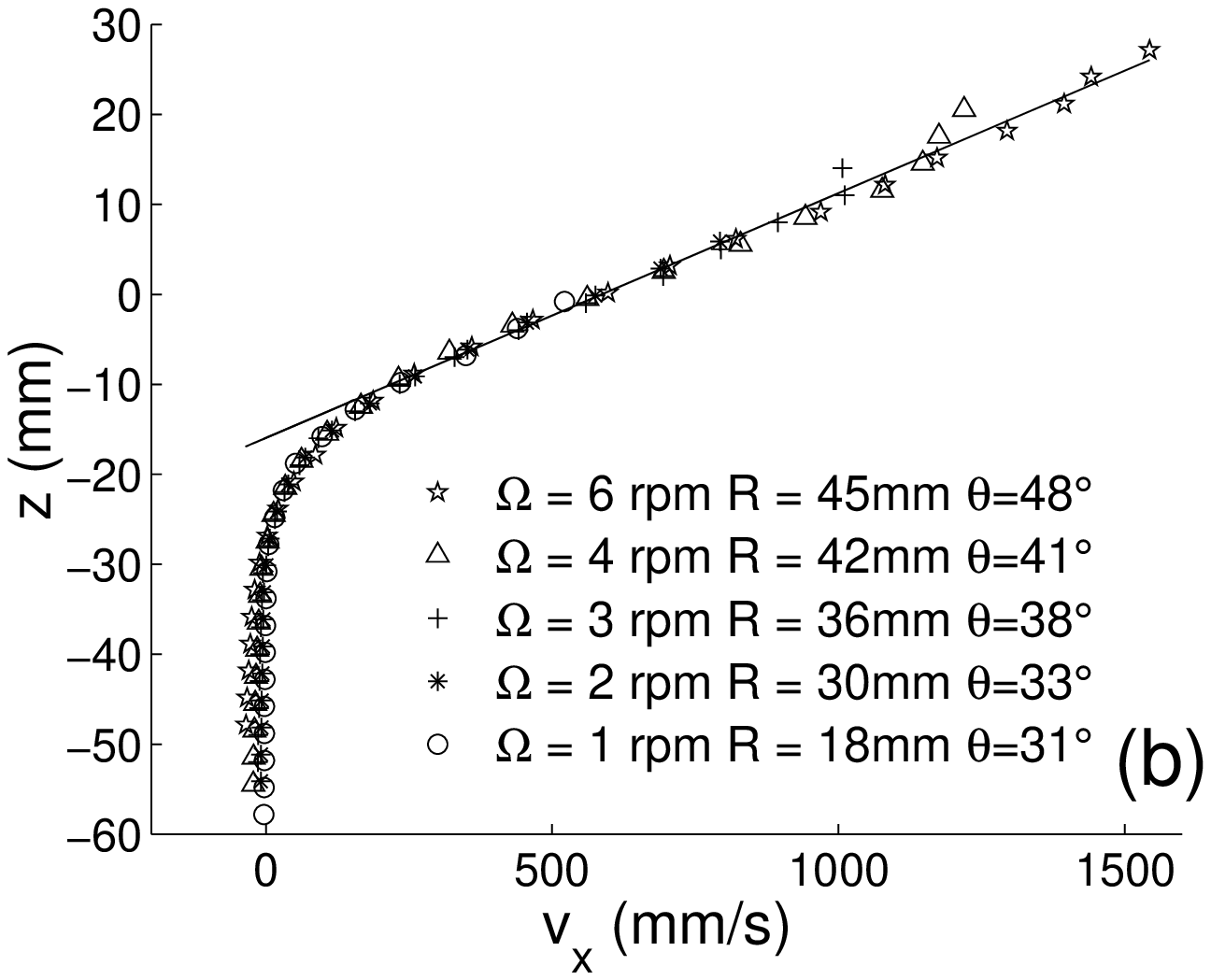, width=\textwidth}
\end{minipage}
\vspace{0.5cm}
\caption{Bonamy, Phys. Fluids}
\label{f.5}
\end{figure}

The rotation speed $\Omega$ has been varied from $1\un{r.p.m.}$ to $6\un{r.p.m.}$ in order to study variations of the velocity profile with $R$ and $\theta$. The corresponding velocity profiles have been represented fig.~\ref{f.5}a. In fig.~\ref{f.5}b, all profiles have been translated along the $z$-axis to make the static/flowing interface coincide. The velocity profile remains linear in the whole flowing layer and the velocity gradient is independent of both $R$ and $\theta$.

\begin{figure}
\centering\epsfig{figure=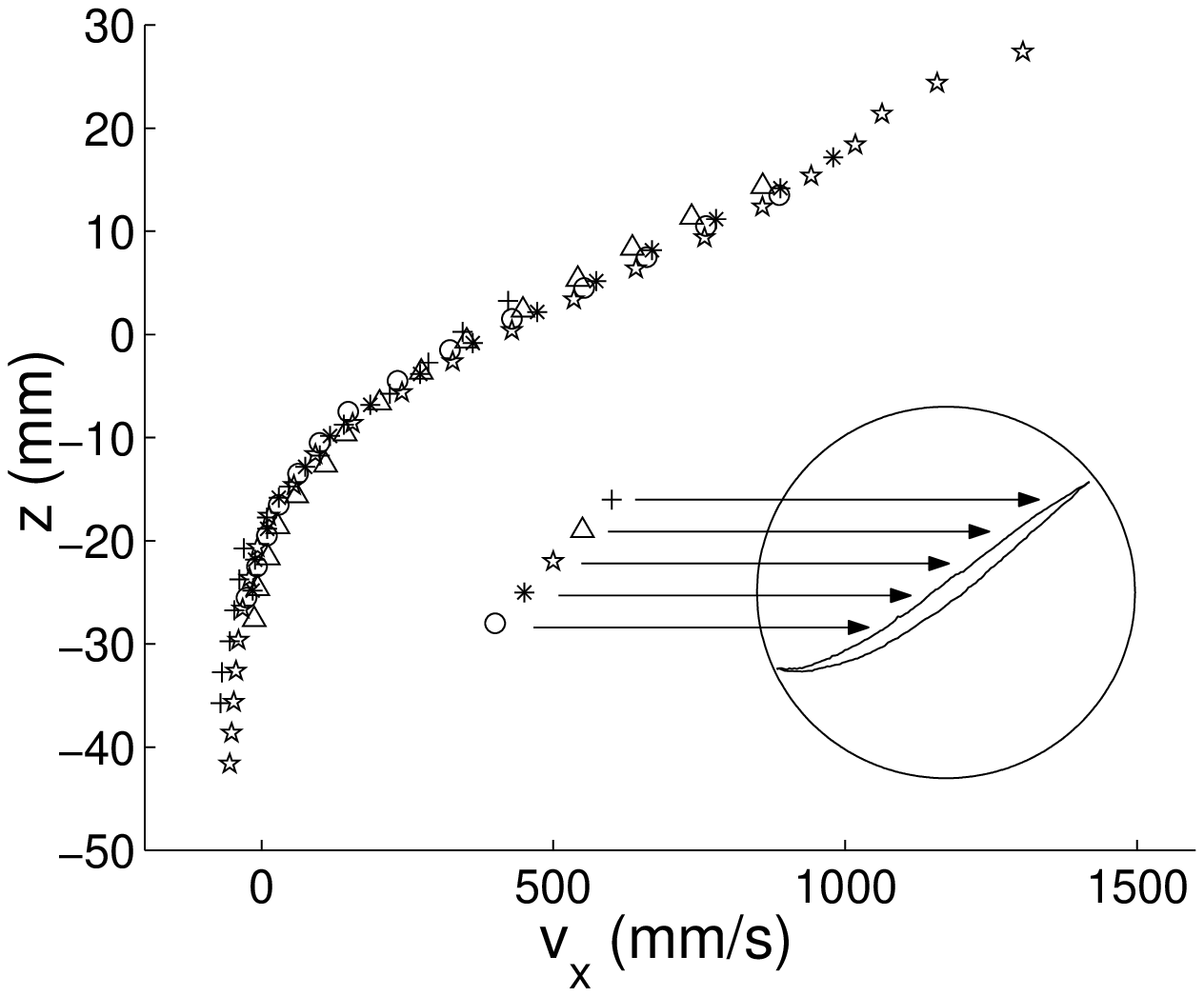, width=0.45\textwidth}
\vspace{0.5cm}
\caption{Bonamy, Phys. Fluids} 
\label{f.5bis}
\end{figure}

Velocity profiles measured at five different points equally distributed along the free surface for $\Omega=6\un{r.p.m.}$ are represented in fig.~\ref{f.5bis}. At these locations, ${\rm d}_{x} R$ and ${\rm d}_{x} \theta$ are no more equal to zero. The thickness of the recorded region has been decreased to minimize these drifts. $\{v_{x}(z)\}$ does not depend on the location in the drum, which means that $\{v_{x}(z)\}$ does not depend on ${\rm d}_{x} R$ and ${\rm d}_{x} \theta$. The velocity profile can be then written inside the flowing layer as:

\begin{equation}
v_{x}(z)=\dot{\gamma} (z+R)
\label{e.10}
\end{equation}

\noindent with $\dot{\gamma}=34\pm 0.5\un{s}^{-1}$, or $\dot{\gamma} \simeq 0.6\sqrt{g/d}$. Such order of magnitude can be easily understood~\cite{rajchenbach00}: when a bead collides with an underlying grain, all the kinetic energy is absorbed by multiple collisions. Then, the balance between potential and kinetic energy on a mean free path of order of $d$ leads to a limiting velocity between two adjacent beads layers equal to $\sqrt{g d}$, and consequently to a velocity gradient of order of $\sqrt{g/d}$. 

The knowledge of the profile $\{v_{x}(z)\}(R,\theta)$ enables us to calculate the first two closure relations:

\begin{equation}
<v_{x}>(R,\theta)=\frac{1}{2} \dot{\gamma} R
\label{e.11}
\end{equation}

\begin{equation}
<{v_{x}}^2>(R,\theta)=\frac{1}{3} \dot{\gamma}^2 R^2
\label{e.12}
\end{equation}

This scaling law is different from the one observed in granular flows down rough inclined plane for 2D packing~\cite{azanza97} and for 3D packing~\cite{pouliquen99}: In these experiments, $<v_{x}>$ was found to scale as $R^{3/2}$. A linear profile with a velocity gradient independent of $R$ has also been observed by Rajchenbach at the surface of a 2D packing in a rotating drum~\cite{rajchenbach00}. This tends to confirm that boundaries conditions can strongly affect the velocity profile and questions the ability of a classical constitutive law relating the shear stress to the strain rate~\cite{mills99,rajchenbach00}. 

Let us note that velocity gradient observed by Rajchenbach in 2D packing scales as $\sqrt{\sin\theta}$ contrary to ours. However, in our series of experiments, $\theta$ varies only from $31^\circ$ up to $48^\circ$, which corresponds to variations of $\sqrt{\sin\theta}$ from $0.72$ to $0.86$. Consequently, the difference between $\dot{\gamma}$ behaving as $\sqrt{\sin\theta}$ and $\dot{\gamma}$ constant does not differ much from the error bars and it is not possible to conclude at this point which of the two scaling is the right one. This will be tested on a more significant range of $\theta$, from $0^\circ$ up to $48^\circ$, in the next section.
 
\section{macroscopic scale: profiles of the flowing phase height and of the static/flowing boundary} \label{S5}

\begin{figure}
\centering
\begin{minipage}[c]{0.45\textwidth}
\centering\epsfig{figure=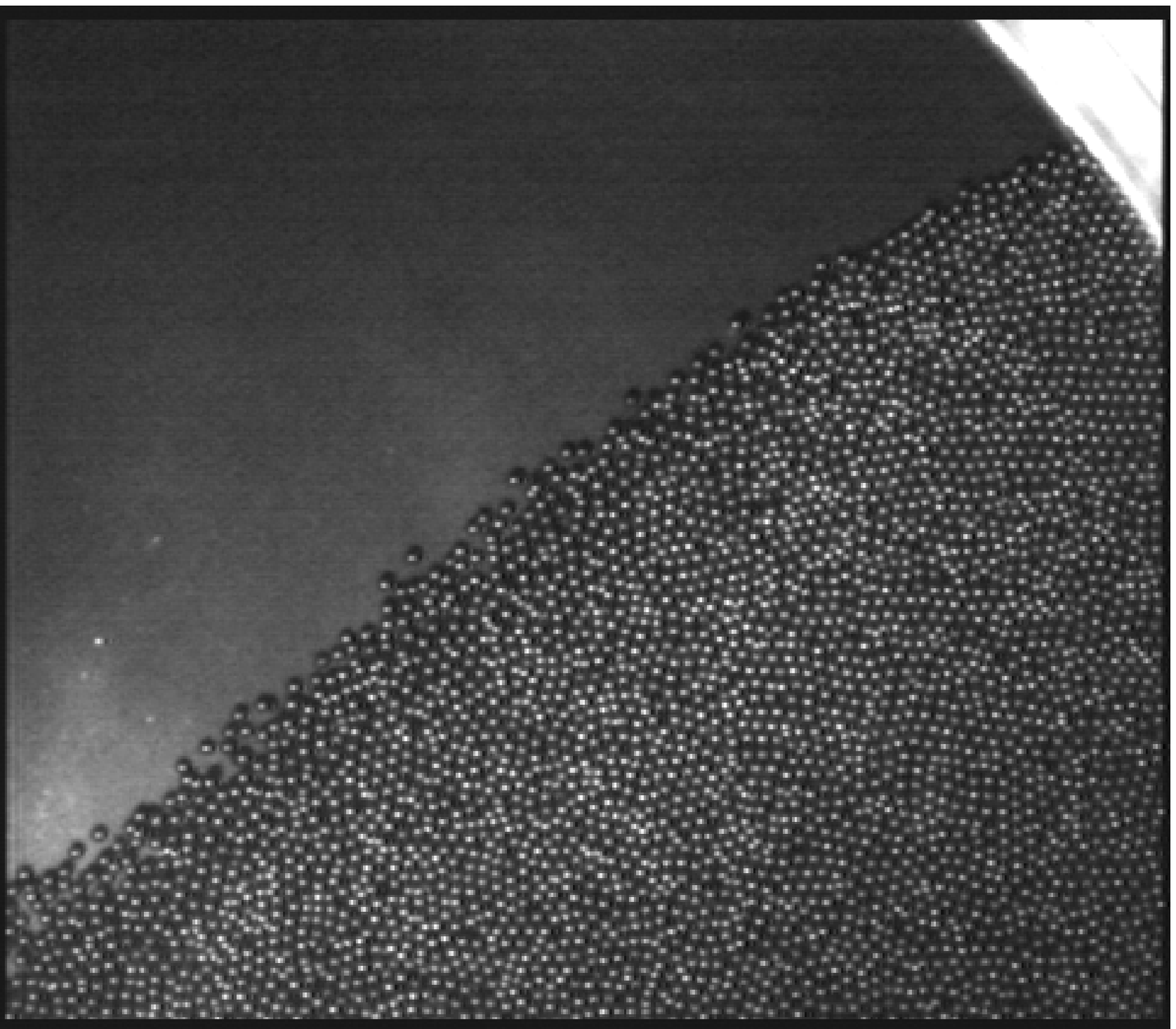, width=\textwidth}
\end{minipage}
\begin{minipage}[c]{0.45\textwidth}
\centering\epsfig{figure=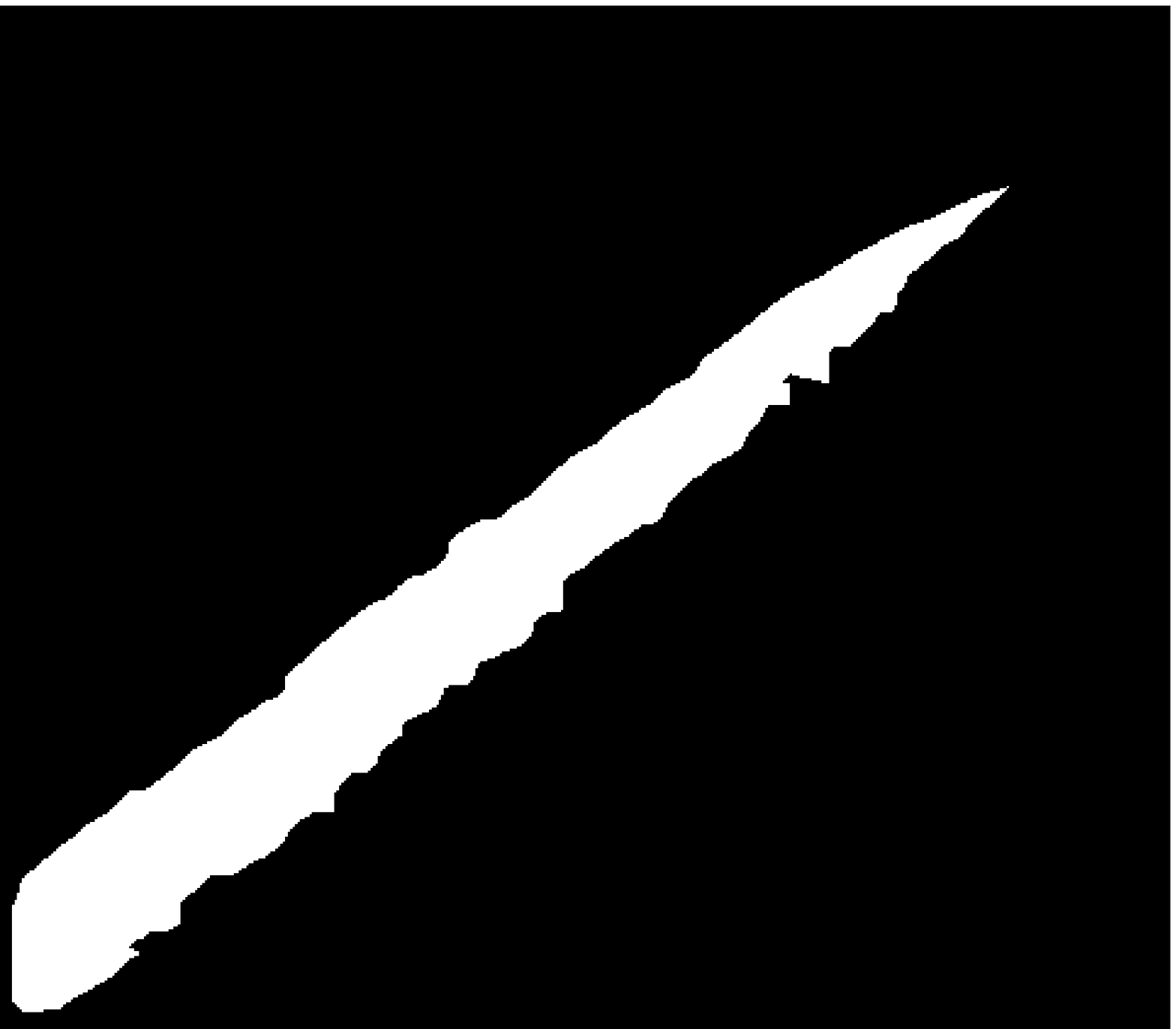, width=\textwidth}
\end{minipage}
\vspace{0.5cm}
\caption{Bonamy, Phys. Fluids}
\label{f.6}
\end{figure}

\begin{figure}
\centering
\begin{minipage}[c]{0.35\textwidth}
\centering\epsfig{figure=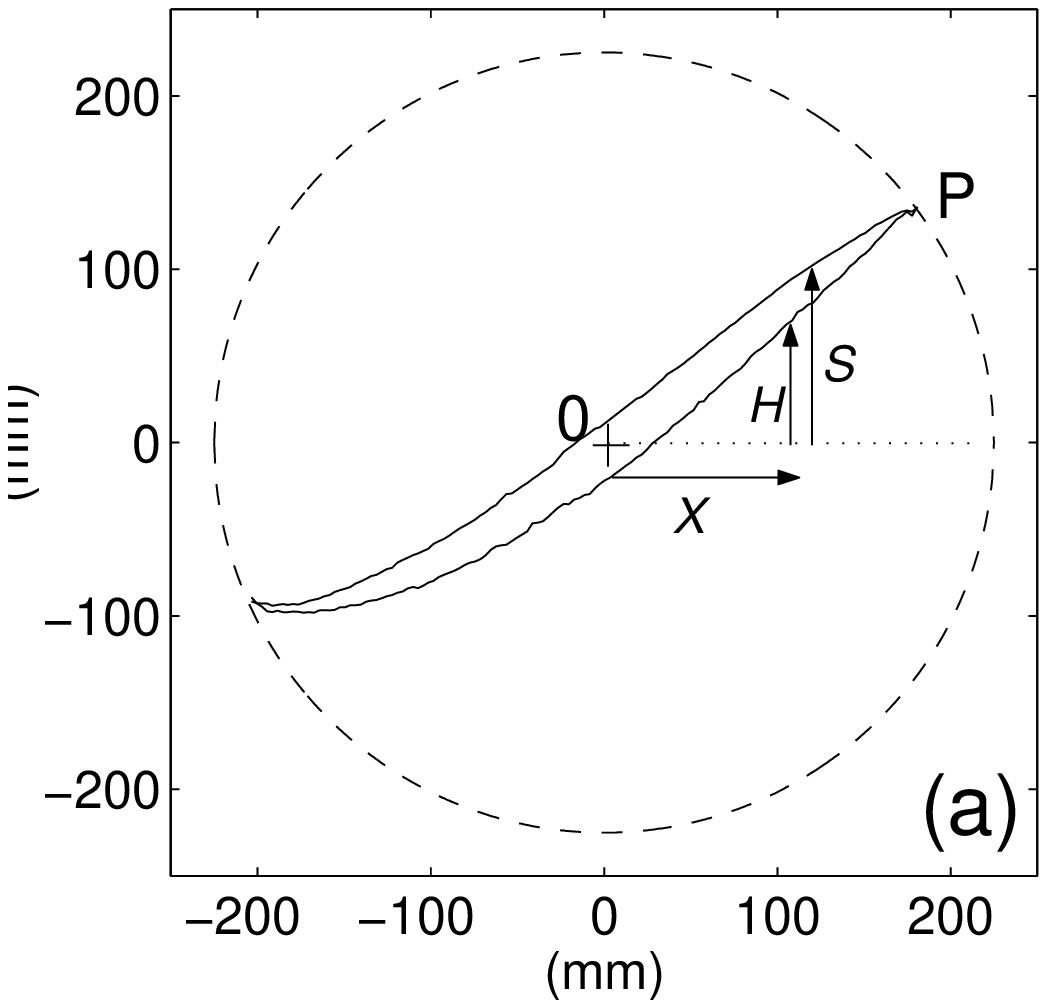, height=6cm}
\end{minipage}
\begin{minipage}[c]{0.55\textwidth}
\centering\epsfig{figure=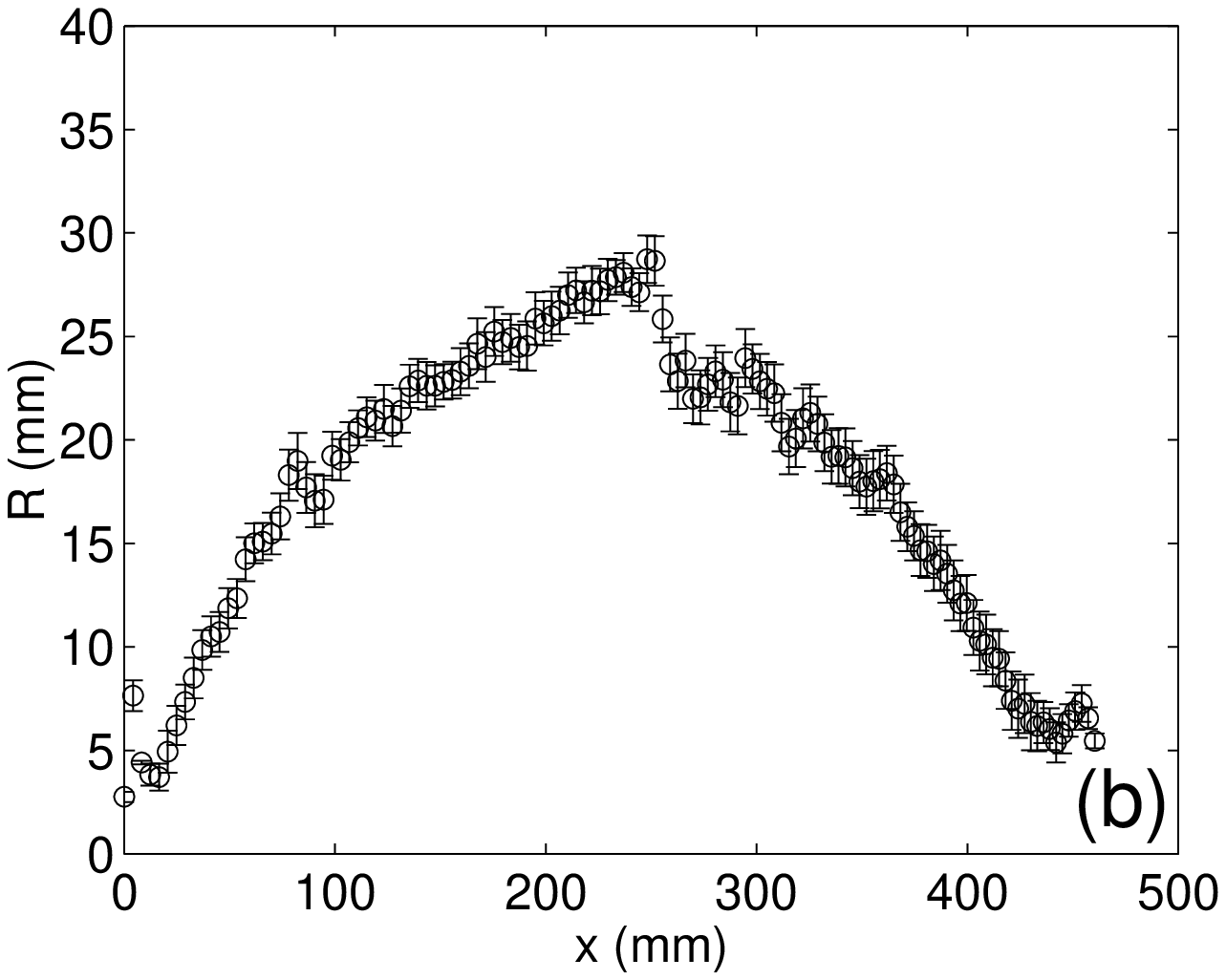, height=5.8cm}
\end{minipage}
\vspace{0.5cm}
\caption{Bonamy, Phys. Fluids}
\label{f.7}
\end{figure}

The determination of the last closure relation $F(R,\theta)$ requires the determination of $R$ and $\theta$ in the whole drum. The experimental protocol is the following: sequences of 512 frames are recorded via a fast camera at a sampling rate $f_s = 250\un{Hz}$ with a pixel resolution of $480 \times 400$. In order to keep a sufficient resolution on the profiles, the rotating drum is visualized by halves, a pixel corresponding to $0.468\un{mm}$. Image processing allows to isolate the flowing layer and both the interface profile $H(X)$ and the free surface profile $S(X)$, expressed in the gravity frame $\gravity$, can be extracted (cf fig.~\ref{f.6}). The fluctuations on each point of these profiles are around $1.5\un{mm}$. The profiles are then averaged over $20$ frames. Consequently, the resulting errors on both $H$ and $S$ are around $0.3\un{mm}$. The curvilinear coordinate $x(X)$, the local angle $\theta(X)$, and the flowing layer thickness $R(X)$ are then deduced using the procedure given in appendix~C. Resulting errors on $R$ is about $1\un{mm}$ (see fig.~\ref{f.7}b). On the upper part of the profile, the local slope is constant on a range whose size decreases when $\Omega$ is increased (see fig.~\ref{f.7}a). One can study on this range ${\mathcal{L}}_{\Omega}$ the variation of the different variables with $R$ for a fixed angle $\overline{\theta}_\Omega$ depending on the rotation speed $\Omega$.

\begin{figure}
\centering
\begin{minipage}[c]{0.45\textwidth}
\centering\epsfig{figure=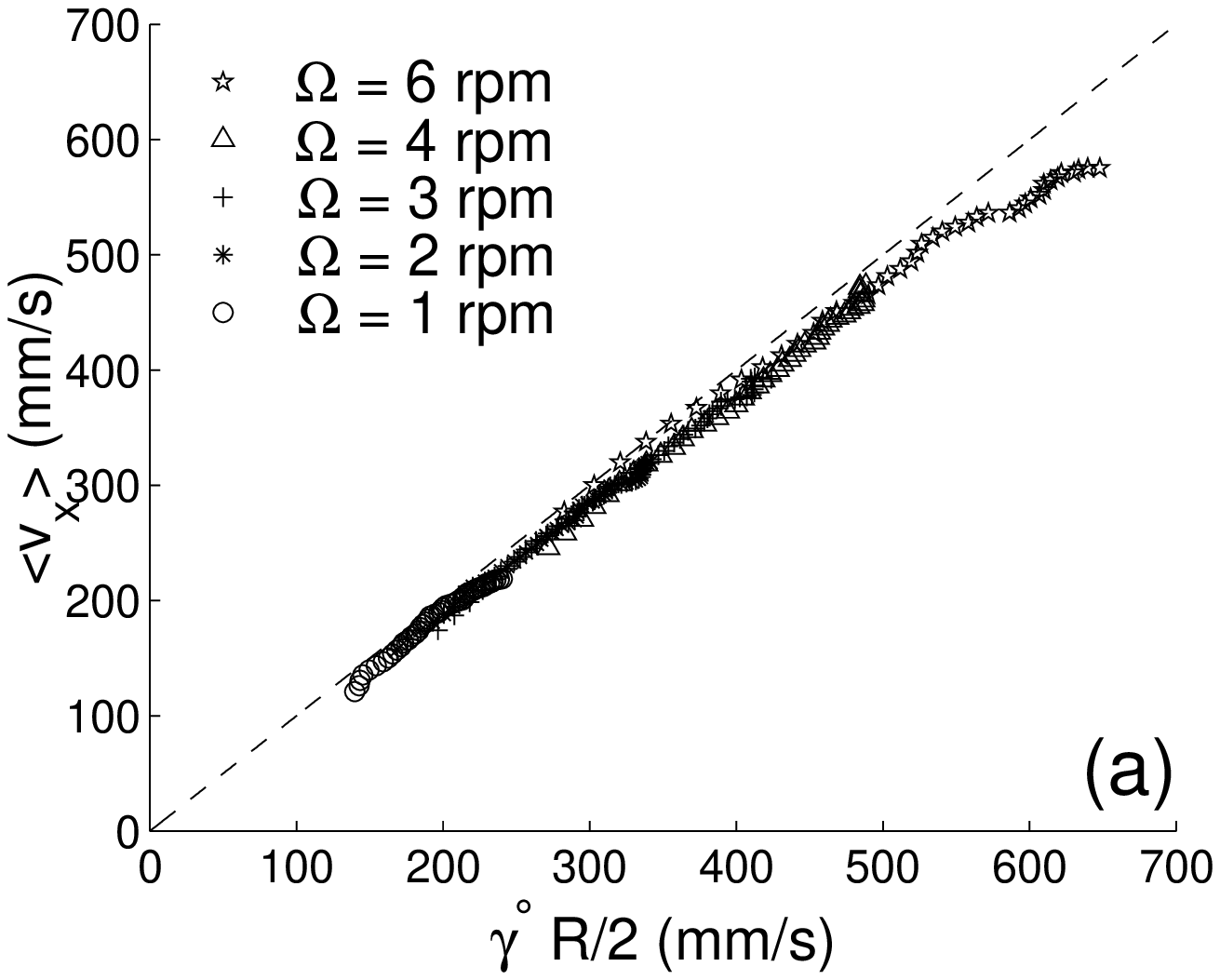, width=\textwidth}
\end{minipage}
\begin{minipage}[c]{0.45\textwidth}
\centering\epsfig{figure=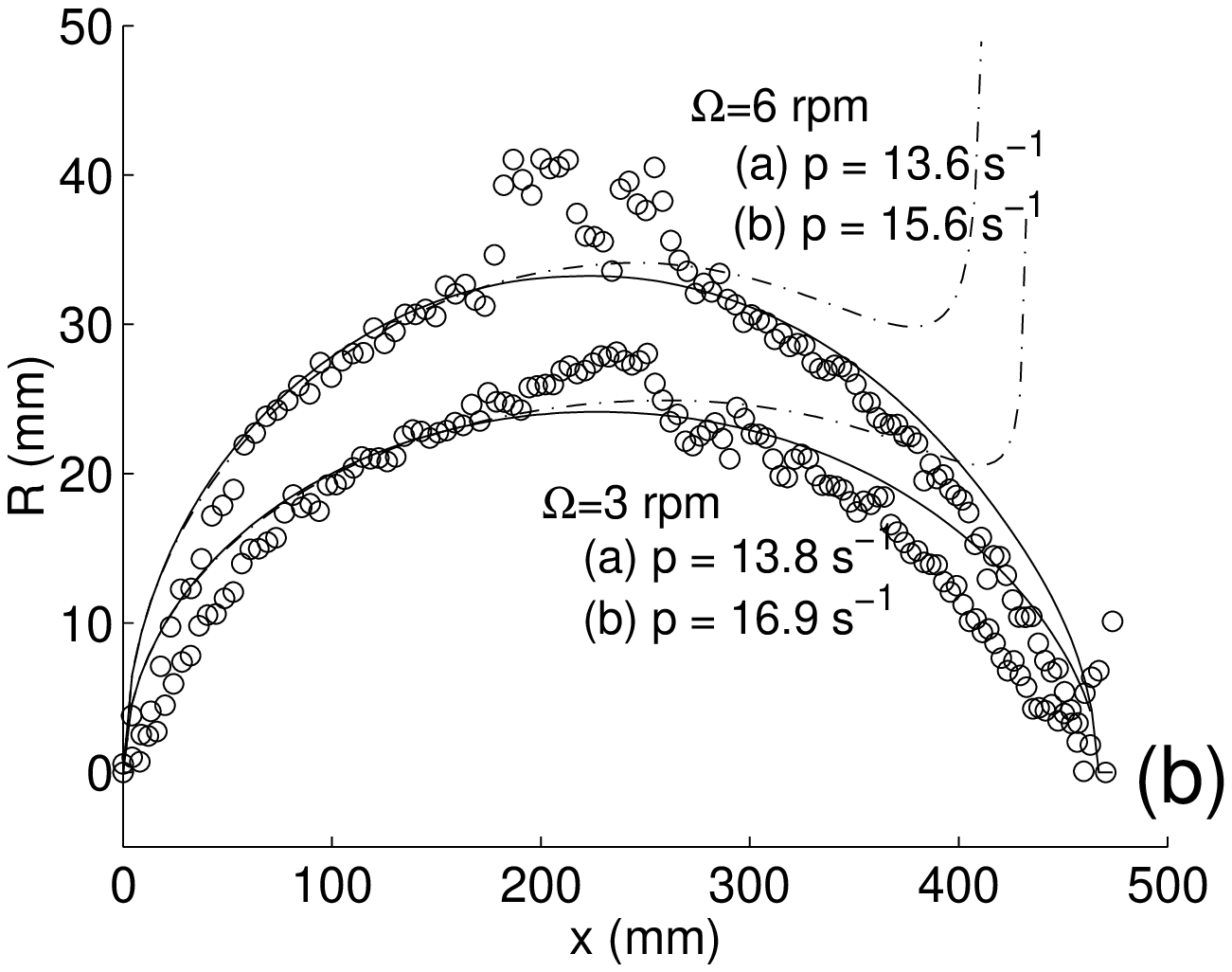, width=\textwidth}
\end{minipage}
\vspace{0.5cm}
\caption{Bonamy, Phys. Fluids}
\label{f.8}
\end{figure}

Let us first confirm the first closure relation (eq.~\ref{e.11}) established at the end of the last section: the drift velocity $<v_{x}>$ can be calculated from the experimental $x(X)$, $H(X)$ and $R(X)$ profiles by integrating numerically eq.~\ref{e.6}. Fig.~\ref{f.8}a  shows the  variation of $<v_{x}>$ vs. $\frac{1}{2}\dot{\gamma}R$ with $\dot{\gamma}=34\un{s}^{-1}$ as suggested in the previous section for different $\overline{\theta}_\Omega$. This confirms the assumption of a linear profile $\{v_{x}(z)\}$ with a velocity gradient independent of $R$.

Let us answer now to the question about the dependence of $\dot{\gamma}$ with $\theta$. To test the different assumptions on this scaling, namely if $\dot{\gamma}$ scales as $\sqrt{\sin\theta}$ or is independent of $\theta$, profiles $H(X)$ and $R(X)$ have to be taken into account in their whole: Typically, for $\Omega=6\un{r.p.m.}$, $\theta(X)$ varies from $0^\circ$ to $48^\circ$. The procedure will be the following: from the profile $H_{exp}(X)$ obtained experimentally, one deduces $X(x)$, $H(x)$, $\theta(x)$ and, by integrating the eq.~\ref{e.6}, the flow rate $\{R<v_{x}>\}(x)$. The theoretical profiles for the flowing thickness $R^{(a)}_{th}(x)$ and $R^{(b)}_{th}(x)$ are then calculated assuming respectively the relations:

\begin{itemize}
\item[(a)] $R<v_{x}>=pR^2$ corresponding to a velocity gradient independent of $\theta$
\item[(b)] $R<v_{x}>=pR^2\sqrt{\sin\theta}$ corresponding to a velocity gradient $\dot{\gamma}$ scaling as $\sqrt{\sin\theta}$ as observed by Rajchenbach~\cite{rajchenbach00}
\end{itemize}

These two profiles are then  compared to the experimental profile $R_{exp}(x)$ (see fig.~\ref{f.8}b). The fit parameter is $p$. The scenario (a) leads to better fits, with a fit parameter $p=14\un{s}^{-1}$ about $20 \%$ smaller than $\dot{\gamma}/2$. This difference may come from the boundaries: the velocity gradient $\dot{\gamma}$ has been determined from the velocity of only the beads observed through the porthole while $p$ is related to the mean velocity of all the beads.   

At this point, one can evaluate the force $F$. Numerical calculation of the second left-hand term of eq.~\ref{e.7} shows that it is negligible compared to the first left-hand term and can thus be neglected.
From eq.~\ref{e.6}, the simplified eq.~\ref{e.7} and the closure relations \ref{e.11} and \ref{e.12}, one finds:

\begin{equation}
\frac{F}{\rho}=\dot{\gamma} R \Omega\Big(H\sin\theta+X\cos\theta\Big)
\label{e.14}
\end{equation}

\begin{figure}
\centering
\begin{minipage}[c]{0.45\textwidth}
\centering\epsfig{figure=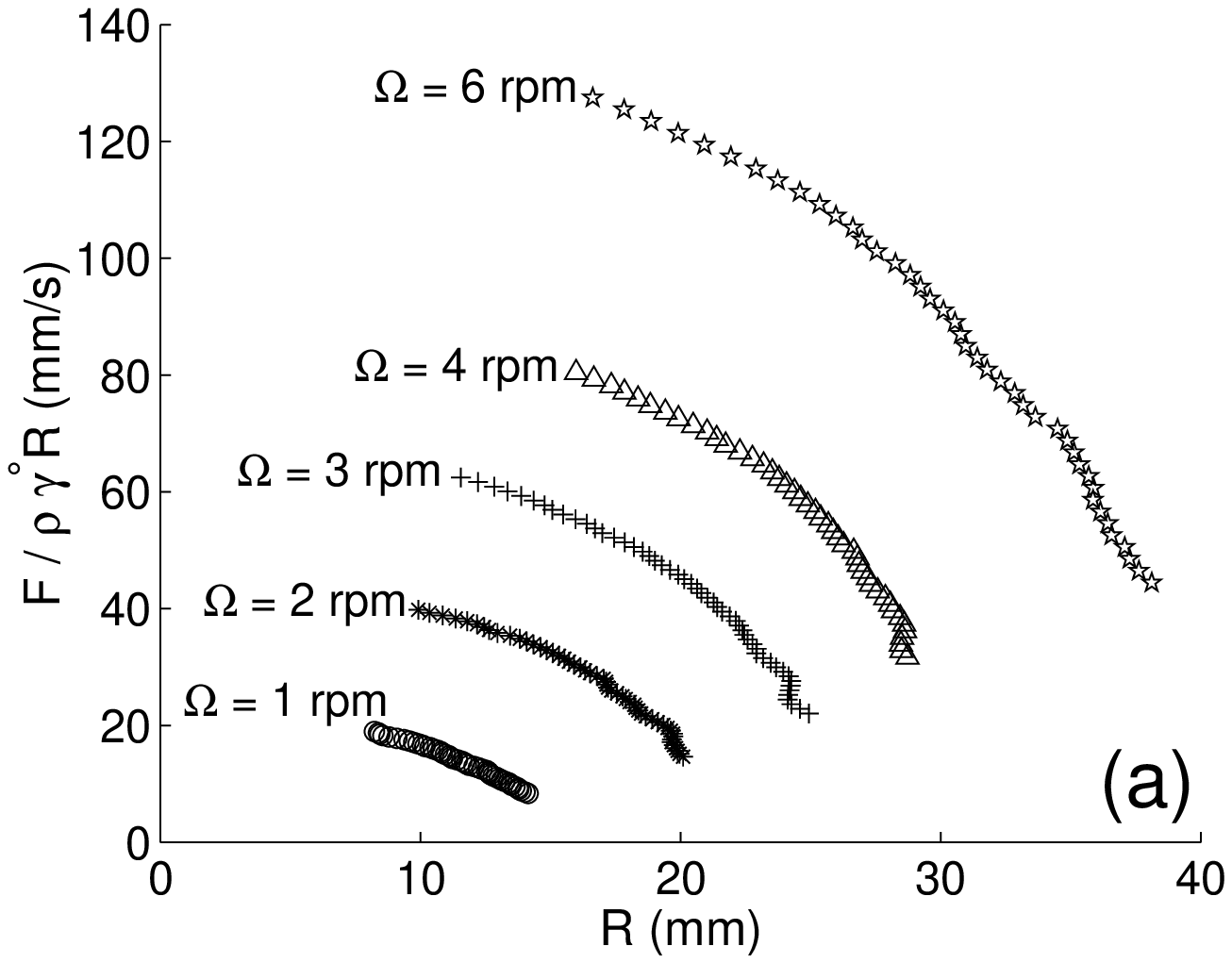, width=\textwidth}
\end{minipage}
\begin{minipage}[c]{0.45\textwidth}
\centering\epsfig{figure=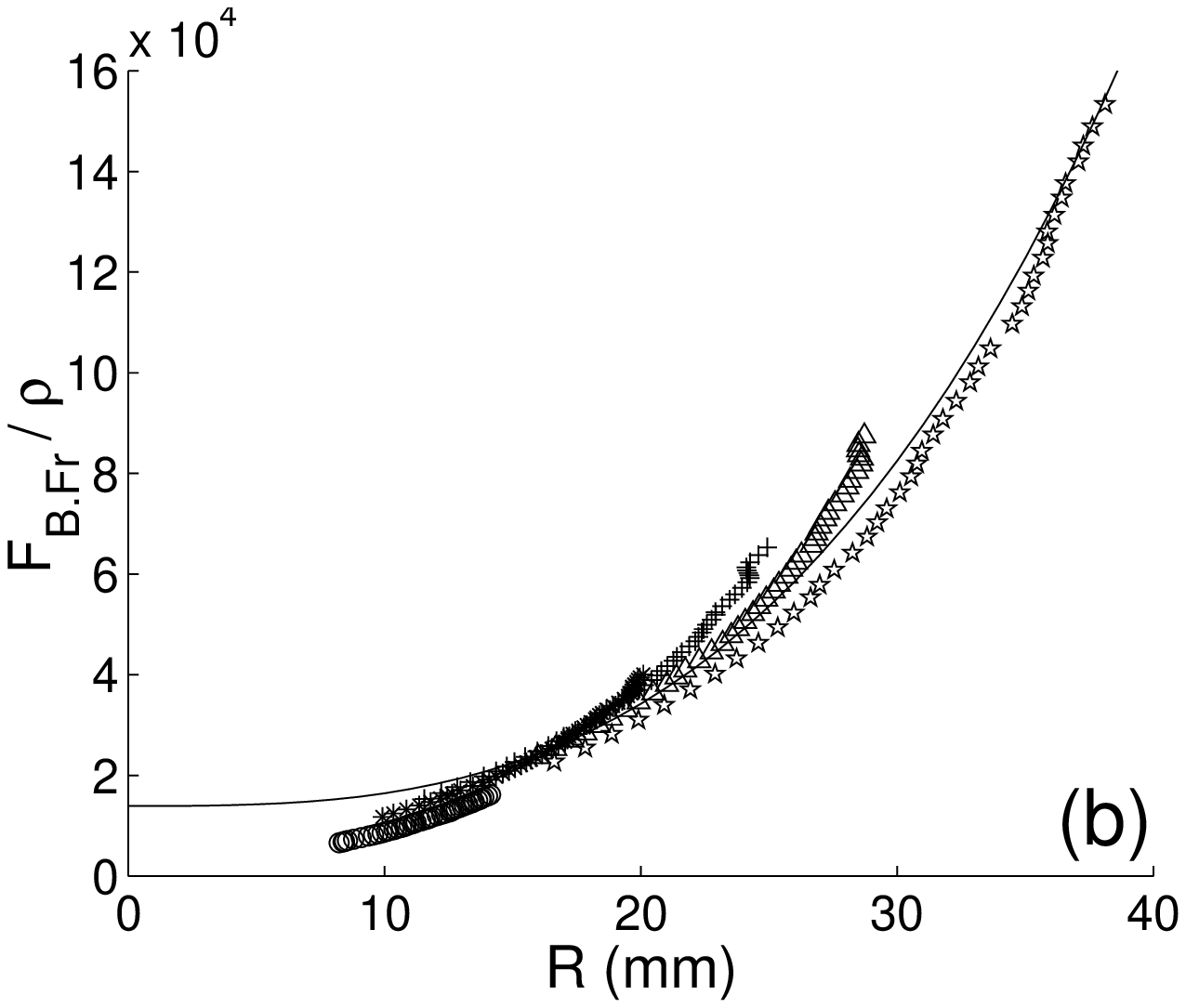, width=\textwidth}
\end{minipage}
\vspace{0.5cm}
\caption{Bonamy, Phys. Fluids}
\label{f.9}
\end{figure}

The last closure relation can then be determined. From the experimental $X(x)$, $H(x)$ and $R(x)$ profiles, one can deduce $F(x)$ using eq.~\ref{e.14}. Fig.~\ref{f.9}a presents the variation of $F/\rho \dot{\gamma} R$ vs. $R$ for different $\theta$. $F/\rho \dot{\gamma} R$ seems to decrease as a parabola with $R$:

\begin{equation}
\frac{F}{\rho \dot{\gamma} R}=F_1(\theta)-F_2(\theta){R}^2
\label{e.15}
\end{equation}

A minimal form of the force $F$ should include:
\begin{itemize}
\item the weight ${F}_W=\rho g R \sin \theta$
\item a pressure term ${F}_P=-k \rho g \frac{\mathrm{d} R}{\mathrm{d} x} \cos \theta$~\cite{pouliquen99bis}. Since $R$ is negligible compared to $D_0$, this term can be neglected.
\item the friction on the static phase: ${F}_{S.fr}=-\mu \rho g R\cos \theta$. A priori, $\mu$ depends on $R$ and $\theta$. Experiments shows however that $\mu$ reach quickly a constant value close to $\tan \theta_r$ independent of $R$ and $\theta$ as soon as $R$ is larger to few beads diameters~\cite{daerr00}. In our case, $\mu$ can be considered as constant.
\item finally, the friction on the lateral boundaries ${F}_{B.fr}=\int_0^{R}\sigma_{B.fr}\mathrm{d} z$.

\end{itemize} 

This last contribution is the one responsible for the decrease of $F/\rho \dot{\gamma} R$ with $R$ as a parabola. It can be obtained directly from the value of $F$ and the first two contributions $F_{W}$ and $F_{S.fr}$. Fig~\ref{f.9}b represents the variation of $F_{B.fr}/ \rho$ vs. $R$ for different $\overline{\theta}_\Omega$. Its variation with $\overline{\theta}_\Omega$ appears to be weak. The form of 
${F}_{B.fr}$ can be understood through the following process: ${F}_{B.fr} = \int_0^{R}\sigma_{B.fr}{\rm d}z$ where $\sigma_{B.fr}(z)$ is the stress exerted by the boundary on the layer between $z$ and $z+{\rm d} z$. To model $\sigma_{fr.B}$, let us apply the following process: $\sigma_{fr.B}=\nu \delta p$ where $\nu$ is the frequency of bead-boundary collision and $\delta p$ the $x$-momentum lost by each collision. As $\rho$ is constant, the mean free path is constant in the layer and $\nu \propto \rho v_{x}(z)$. Each collision with the boundary is inelastic. Consequently, $\delta p \propto v_{x}$. Finally one can write $\sigma_{fr.B}=\eta \rho {v_{x}}^2$ with $\eta$ constant and, after integration over the width of the flowing layer:

\begin{equation}
{F}_{fr.B}=-\rho \frac{\eta \dot{\gamma}^2}{3}{R}^3 
\label{e.16}
\end{equation}

This closure relation indeed allows one to reproduce the experimental $F_{B.fr}(R)$ profiles obtained for different $\overline{\theta}_\Omega$ (cf. fig~\ref{f.9}b). In our experiment, $\eta \simeq 10 \un{m}^{-1}$. The order of magnitude of $\eta$ can be understood in the following way: the momentum lost by each collision is of order $(1-e)$ where $e$ is the restitution coefficient. For our beads, $e\simeq 0.95$. The mean free path is of the same order of the bead diameter $d$. Consequently, one expects $\eta \simeq (1-e)/d \simeq 10 \un{m}^{-1}$.

The last closure relation that follows from the above discussion thus reads:

\begin{equation}
F=\rho g R \sin \theta -\mu \rho g R \cos \theta-\frac{1}{3}\eta \dot{\gamma}^2 {R}^3
\label{e.17}
\end{equation}

\section{discussion} \label{S6}

Using the closing relations eq.~\ref{e.11}, eq.~\ref{e.12} and eq.~\ref{e.17}, the conservation equations (\ref{e.1},\ref{e.2}) become:
    
\begin{equation}
\frac{\partial R}{\partial t}+V\frac{\partial R}{\partial x}=\Gamma
\label{e.18}
\end{equation}

\begin{equation}
\frac{\partial H}{\partial t}=-\Gamma
\label{e.19}
\end{equation}

\noindent where:

\begin{equation}
V=2<v_x>=\dot{\gamma}R
\label{e.20}
\end{equation}

\noindent and: 

\begin{equation}
\Gamma=\frac{F}{\rho \dot{\gamma} R}=\frac{g}{\dot{\gamma}}(\sin\theta -\mu\cos\theta) -\frac{1}{3}\eta \dot{\gamma} R^2
\label{e.21}
\end{equation}

\noindent This system of equations contains three parameters: the velocity gradient $\dot{\gamma}$, the friction coefficient $\mu$ and a third parameter $\eta$ characterizing the friction on lateral boundaries. These three parameters depend on microscopic features of the beads such as their diameter, their coefficient of restitution and their roughness and of the gap of the drum. It can be compared to the different existent continuous models proposed in~\cite{douady99,khakhar01,bouchaud94,boutreux98,aradian98}. 

These two equations are very close to the one proposed by Douady {\em et al.}~\cite{douady99}. They have supposed a linear velocity profile with a constant velocity gradient as observed in our experiment. Moreover, for large $R$, variation of the force $F$ with $R$ and $\theta$ is qualitatively similar to ours, but they attribute this variation to the friction of the flowing layer to static grains instead of invoking friction effects on the lateral boundaries as we have done.

 Khakhar {\em et al.}~\cite{khakhar01} have proposed a derivation of depth averaged conservation equations very close to that of \cite{douady99}. They also have assumed a linear velocity, but with a gradient dependent of $\theta$. As for the closing relations proposed by \cite{douady99}, variation of $F$ with $R$ is attributed to a dependency of the shear stress at the solid/flowing interface with $R$ rather than to boundary effects.  

Another model has also been proposed a few years ago by Bouchaud {\em et al}~\cite{bouchaud94}. The original model, devoted  to describe avalanche triggering, deals with thin moving layers of $R\leq d$ and can not be compared to ours. But this model has later been extended to describe thick surface flows by Boutreux {\em et al.}~\cite{boutreux98}.
These models assume a constant advection velocity $V$, small variations of $\theta$ around $\theta_r$, and relate the coupling term $\Gamma$ to the effect of collisions between the moving grains on the static bed. For thin flowing layers~\cite{bouchaud94}, all the moving grains are assumed to be in contact with the static bed and to have the same probability to interact with it: $\Gamma = a R(\theta-\theta_r)$ whith $a$ constant. For thick granular flows~\cite{boutreux98}, only grains belonging to the first layers can interact with the static bed. The influence of the others is shielded by these first layers. $\Gamma$ is then more naturally assumed to be independent of $R$: $\Gamma = a (\theta-\theta_r)$ with $a$ constant. This last form is compatible with our form of $\Gamma$ expanded to first order in $\theta-\theta_r$. The constant $a$ can be evaluated~\cite{douady99,khakhar01}: $a=g\cos\theta_r/\dot{\gamma}$. Thanks to their simplicity, these two last model allow analytical exact solutions~\cite{mahadevan99,dorogovtsev99,thorsten00}. But the assumption of a constant velocity $V$ prevents them for describing quantitatively avalanches front~\cite{clade01} or avalanches amplitudes~\cite{aradian98}.

\section{conclusion}

Depth averaged conservation equations have been written for granular surface flows. Their applications to the study of steady surface flows in a rotating drum allow one to find experimentally the closure relations of these equations from measurements of the velocity profile in the flowing layer at the center of the drum and from the flowing layer thickness $R$ and the interface slope $\theta$ profiles. These measurements have been performed for a quasi 2D packing of steel beads. The velocity profile has been found linear with a gradient $\dot{\gamma}$ independent of both $R$ and $\theta$. The closure relations have been deduced: the mean drift velocity $<v_{x}>$ and the mean kinetic energy $<{v_x}^2>$ are given respectively by $<v_x>=\frac{1}{2}\dot{\gamma}R$ and $R<{v_x}^2>=\frac{1}{3}\dot{\gamma}^2 R^2$. The last relation relating the force $F$ acting on the flowing layer to $R$ and $\theta$ has also been explicitly given and reveals the importance of lateral boundaries. Two parameters, namely the friction coefficient $\mu$ of the flowing layer on the static bed (close to $\tan\theta_r$) and a coefficient $\eta$ characterizing the friction on lateral boundaries are needed to describe $F$. Finally, the closed equations have been compared to existent continous models. Their differences have been discussed: a new term whose form is given should be included to take into account lateral boundaries effects. The precise study of the dependence of the three parameters $\dot{\gamma}$, $\mu$ and $\eta$  with the beads diameters, the beads coefficient of restitution, the beads surface roughness and the cell gap represents interesting topic for a future investigation.

\acknowledgments

We  thanks J.-Ph Bouchaud for a critical reading of this manuscript.
We are grateful to B. Faucherand and M. Planelles for their participation to the data collection. We also acknowledge discussions with M. Bonetti, P. Claudin, A. Daerr, S. Douady, P. Evesque, D. Khakhar and O. Pouliquen. We thank C. Gasquet and P. Meininger for technical support.

\section{appendix}

\subsection{Determination of the volume fraction profile}

\begin{figure}
\centering
\epsfig{figure=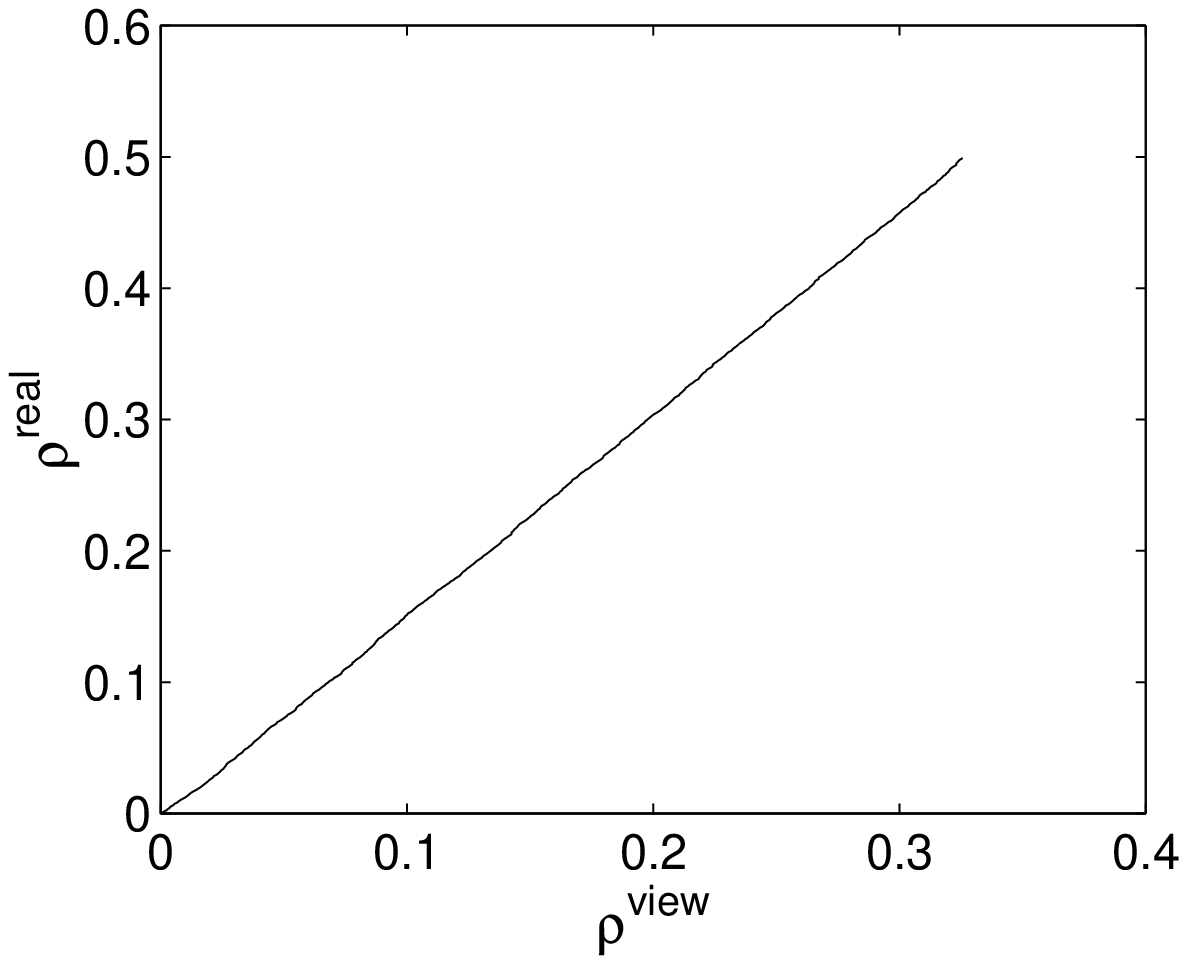, , width=0.45\textwidth}
\vspace{0.5cm}
\caption{Bonamy, Phys. Fluids}
\label{f.10}
\end{figure}

3D disordered packing of different volume fractions confined between two plates have been simulated. Numerical packing have been constructed by dropping randomly beads in a cell of gap $e$. Beads are randomly located at a given altitude. Their altitudes are then minimized under the two following constrains: the beads should stay in the cell and the distance between their center of mass and the center of mass of beads already in the cell should be superior to $d$. Confined packing of volume fraction $0.5$ can then be obtained. Confined packing of lower volume fraction are then obtained either by contracting beads and cell dimensions without changing the position of beads center of mass or by removing randomly a given number of beads. One can then determine the beads actually observed through the porthole, {\em i.e.} the beads whose center of mass is not hidden by the the presence of the other beads closer to the porthole. Both real volume fraction $\rho^{real}$ and the volume fraction $\rho^{view}$ evaluated from the observed beads are calculated (see fig.~\ref{f.10}). $\rho^{real}$ varies linearly with $\rho^{view}$. For $d=3\un{mm}$ and $e=7\un{mm}$:

\begin{equation}
\rho^{real} =1.61 \rho^{view}
\end{equation} 

Direct experimental measurements of $\rho^{real}$ and $\rho^{view}$ just after the filling the drum, {\em i.e.} for $\rho^{real}=0.57$ (see section III), confirm the value of the ratio $\rho^{real}/\rho^{view}$. 

\subsection{Determination of the velocity of beads}

The processing to determine beads velocity is the following: each bead is tracked over ten successive frames. As soon as the displacement between two successive frames is larger than $d$, there is an ambiguity and the bead is considered as lost. Then, only beads tracked over at least five successive frames are kept. They represent more than $99\%$ of all the observed beads. The largest accessible velocity is thus $f_s d/2=1500\un{mm/s}$ and the error on these velocities is about $10\un{mm/s}$. The local angle $\theta$ of the surface flow with the horizontal axis is then determined from the mean velocity of the flow calculated by averaging the velocity of all beads in all frames. All frames are then rotated by $\theta$ and divided into horizontal layers one bead diameter wide. The velocity profile is then calculated by averaging the velocity of the beads in each layer on $200$ frames. The errors made on the velocity profile is then mainly dominated by the statistical processing of the data. It is less than $20\un{mm/s}$ for a $99\%$ confidence interval.

\subsection{Determination of $x$, $\theta$ and $R$}

From experimental $H(X)$ and $S(X)$ profiles, $x(X)$ is calculated via $x=\int_{X}^{X_P}{\sqrt{1+\left(\frac{{\rm d} H}{{\rm d} X}(y)\right)^2}}{\rm d}y$ where $P$ is the upper boundary of the flowing layer ($H(X_P)=S(X_P)$). The local angle $\theta(X)$ is defined as: $\theta={\rm atan}(\ud H / \ud X)$. The procedure  to determine $R(x)$ is the following: for a given $x$, one first finds the corresponding static/flowing boundary point $\mathrm{P}_H$ of coordinate $(X(x),H(x))$ in gravity frame. Then one looks for the point $\mathrm{P}_S$ of coordinate $(X_S,S(x_S))$ at the free surface so that $\overrightarrow{\mathrm{P}_S \mathrm{P}_H}$ be orthogonal to $\overrightarrow{e}_x$ (numerically, this consists of minimising $\Big|(X(x)-X_S)\cos\theta(x)+(H(x)-S(X_S))\sin\theta(x)\Big|$ with respect to $X_S$). $R(x)$ is then defined as $|\mathrm{P}_S \mathrm{P}_H|$.

%
%

\newpage

\begin{itemize}

\item Fig. 1: steady surface flow in a drum at rotation speed $\Omega$. A thin cascading layer of thickness $R$ (in grey) flows on a static bed whose local slope with the horizontal is $\theta$. The transport theorem is applied on the volume limited by the vertical slices $\Sigma (x)$ and $\Sigma (x+\ud x)$ and resulting equations are written in the curvilinear frame $(\vec{e}_x,\vec{e}_z)$. The frame $(\vec{e}_X,\vec{e}_Z)$ is the gravity frame located at the center of the drum ${\rm O}$. The solid rotation at a point $\mathrm{M}$ located at the solid/flowing interface is $\Omega r \vec{n}$ where $r$ is the distance between $\mathrm{O}$ and $\mathrm{M}$ and $\vec{n}$ the unit vector perpendicular to $\overrightarrow{\mathrm{OM}}$.

\item Fig. 2: schematic drawing of the experimental set-up

\item Fig. 3: determination of beads velocities. Left: one of the 200 raw frames for $\Omega=3\un{r.p.m.}$. Frames are binarised and beads position is then defined as the center of mass of white areas. Right: corresponding calculated velocity field. 

\item Fig. 4: left: solid fraction profile (averaged over 1000 frames). Right: velocity profile (averaged over 200 frames) at the center of the rotating drum for $\Omega=6 \un{r.p.m.}$. $v_{x}$ is linear with $z$ in the flowing layer. Inset: $v_{x}/v_{x}|_{z=0}$ vs. the depth showing the exponential creep deformation in the static phase. The straigth line corresponds to an exponential fit. In both figures, the errorbars correspond to a $99\%$ confidence interval.

\item Fig. 5: left: velocity profile at the center of the drum for different $\Omega$: for all these $(R,\theta)$ couples, $v_{x}$ is linear with $z$. Right: all these profiles have been translated along the $z$ axis to make their static/flowing interface coincide. The velocity gradient $\dot{\gamma}$ is independent of both $R$ and $\theta$.

\item Fig. 6: velocity profile for $\Omega=6\un{r.p.m.}$ at five different locations equally distributed along the free surface. The velocity gradient depends weakly on the location in the drum. All these profiles have been tranlated along the $z$ axis to make their static/flowing interface coincide. The velocity gradient $\dot{\gamma}$ is consequently independent of both $\mathrm{d}_x \, R$ and $\mathrm{d}_x \, \theta$.

\item Fig.7: frame processing. Left: one of the $512$ raw frames ($\Omega=3\un{r.p.m.}$). Frames are subtracted 2 by 2, binarised and gathered by bursts of $25$ frames. The resulting frame is smoothed via a dilation-erosion filter. Right: one of the 20 resulting frames showing the flowing layer.

\item Fig. 8: typical profiles obtained for $\Omega=3\un{r.p.m.}$. Left : profiles $H(X)$ and $S(X)$. Dashed circle correspond to rotating drum limits. One can see a large range ${\mathcal{L}}_\Omega$ in the right part of the drum where the layer thickness varies while the static/flowing boundary slope almost does not change. Right : profile $R(x)$.

\item Fig. 9: left: mean drift velocity $<v_{x}>$ vs. $\frac{1}{2}\dot{\gamma}R$ on the range ${\mathcal{L}}_{\Omega}$. Right: test of the relation $v={p}{R}$ in the whole drum. The theoretical profiles ${R_{th}}^{(a)}(x)$ and ${R_{th}}^{(b)}(x)$ are calculated from the experimental profile $H_{exp}(X)$ using the assumptions $(a)$ $v={p}{R}$ (solid line) or $(b)$ $v={p}{R}\sqrt{\sin\theta}$ (dashed line) (see text for details on the procedure). They are then compared to the experimental $R_{exp}(x)$ profile. In all case, $p$ is chosen to minimize $|R_{th}-R_{exp}|$.

\item Fig. 10: left : variation of $F/\rho \dot{\gamma}  R$ vs. $R$ for different $\Omega$. Right: friction force on the lateral plane ${F}_{B.fr}/ \rho$ vs. $R$. The solid line corresponds to eq.\ref{e.16} (translated vertically of a constant value.): $\eta \simeq 10\un{m}^{-1}$.

\item Fig. 11: variation of the real volume fraction profile $\rho^{real}$ vs. the volume fraction profile $\rho^{view}$ taking into account only the beads effectively seen through the porthole. Data come from numerical simulations.

\end{itemize}

\newpage

\newpage

\newpage

\newpage

\newpage

\newpage

\newpage

\newpage

\newpage

\newpage

\newpage

%


\begin{thebibliography}{99}

\bibitem{jaeger96}
H.M. Jaeger, S.R. Nagel and R.P. Behringer, \lq\lq~Granular solids, liquids and gases~\rq\rq, Rev. Mod. Phys. {\bf 68}, 1259 (1996).

\bibitem{mills99}
P. Mills, D. Loggia and M. Texier, \lq\lq~Model for a stationary dense granular flow along an inclined wall~\rq\rq, Europhys. Lett., {\bf 45}, 733 (1999).

\bibitem{venant50}
A.J.C. de Barre Saint-Venant, \lq\lq~M\'emoires sur des formules nouvelles pour la solution des probl\`emes relatifs aux eaux courantes~\rq\rq, C. R. Acad. Sci. Paris, {\bf 31}, 283 (1850).

\bibitem{savage89}
S. Savage and K. Hutter, \lq\lq~The motion of a finite mass of granular material down a rough incline~\rq\rq, J. Fluid Mech., {\bf 199}, 177 (1989).

\bibitem{pouliquen99bis}
O. Pouliquen \lq\lq~On the shape of granular fronts down rough inclined planes~\rq\rq,  Phys. Fluids, {\bf 11}, 1956 (1999).

\bibitem{khakhar97}
D.V. Khakhar, J.J. McCarthy, T. Shinbrot and J.M. Ottino, \lq\lq~Tranverse flow and mixing of granular materials in a rotating cylinder~\rq\rq, Phys. Fluids, {\bf 9}, 31 (1997).

\bibitem{elperin98}
T. Elperin and A. Vikhansky, \lq\lq~Granular flow in a rotating cylindrical drum~\rq\rq, Europhys. Lett., {\bf 42}, 619 (1998).

\bibitem{douady99}
S. Douady, B. Andreotti and A. Daerr, \lq\lq~On granular surface flow equations~\rq\rq, Eur. Phys. J. B, {\bf 11}, 131 (1999).

\bibitem{khakhar01}
D. V. Khakhar, A. V. Orpe, P. Andresen and J. M. Ottino, \lq\lq~Surface flow of granular materials: model and experiments in heap formation~\rq\rq, J. Fluid Mech., {\bf 441}, 255 (2001).

\bibitem{bouchaud94}
J.-Ph Bouchaud, M. Cates, J.R. Prakash and S.F. Edwards, \lq\lq~A model for the dynamics of sandpile surfaces~\rq\rq, J. Phys. France I, {\bf 4}, 1383(1994).

\bibitem{boutreux98}
T. Boutreux, E. Raphael and P.-G. de Gennes, \lq\lq~Surface flow of granular materials: a modified picture for thick avalanches~\rq\rq, Phys. Rev. E, {\bf 58}, 4692 (1998).

\bibitem{aradian98}
A. Aradian, E. Raphael and P.-G. de Gennes, \lq\lq~Thick surface flows of granular materials: the effect of the velocity profile on the avalanche amplitude~\rq\rq, Phys. Rev. E, {\bf 60}, 2009 (1998).

\bibitem{aranson00}
I.S. Aranson and L.S. Tsimring, \lq\lq~Continuous description of avalanches in granular media~\rq\rq, Phys. Rev. E, {\bf 64}, 020301 (2001).

\bibitem{daerr99}
A. Daerr and S. Douady, \lq\lq~Two types of avalanches behaviour in granular media~\rq\rq, Nature, {\bf 399}, 241-243 (1999).

\bibitem{evesque88}
P. Evesque and J. Rajchenbach, \lq\lq~Characterization of glass bead avalanches by using the technique of a rotating cylinder~\rq\rq, C. R. Acad. Sci. Paris, {\bf 307}, 223 (1988).

\bibitem{nagel89}
H.M. Jaeger, C.H. Liu and S.R. Nagel, \lq\lq~Relaxation at the Angle of Repose~\rq\rq, Phys. Rev. Lett., {\bf 62}, 40 (1989).

\bibitem{rajchenbach00}
J. Rajchenbach, \lq\lq~Granular flows~\rq\rq, Adv. Phys., {\bf 49}, 229 (2000).

\bibitem{nakagawa93}
M. Nakagawa S.A. Altobelli, A. Caprihan, E. Fukushima and E.-K. Jeong, \lq\lq~Non Invasive measurements of granular flows by magnetic resonance imaging~\rq\rq, Exp. Fluids, {\bf 16}, 54 (1993).

\bibitem{komatsu01}
T. S. Komatsu, S. Inagaki, N. Nakagawa and S. Nasuno, \lq\lq~Creep motion in a granular pile exhibiting steady surface flow~\rq\rq, Phys. Rev. Lett., {\bf 86}, 1757 (2001).

\bibitem{azanza97}
E. Azanza, \lq\lq Ecoulements granulaires bidimmensionnels sur un plan inclin\'e \rq\rq, PhD thesis, \'Ecole National des Ponts et Chauss\'ees, Noisy le Grand, France (1997)

\bibitem{pouliquen99}
O. Pouliquen, \lq\lq~Scaling laws in granular flows down rough inclined plane~\rq\rq, Phys. Fluids, {\bf 11}, 542 (1999).

\bibitem{daerr00}
A. Daerr, \lq\lq Dynamique des Avalanches \rq\rq, PhD thesis, \'Ecole Normale Sup\'erieure, Paris, France (2000)

\bibitem{mahadevan99}
L. Mahadevan and Y. Pomeau, \lq\lq~Propagating fronts on sandpile surfaces~\rq\rq, Europhys. Lett., {\bf 46}, 595 (1999).

\bibitem{dorogovtsev99}
S. N. Dorogovtsev and J. F. F. Mendes, \lq\lq~How sandpiles spill: sandpiles problems in a thick flow regim~\rq\rq, Phys. Rev. Lett., {\bf 83}, 2946 (1999).

\bibitem{thorsten00}
E. Thorsten, P. Claudin and J. P. Bouchaud, \lq\lq~Exact solutions of a model for granular avalanches~\rq\rq, Europhys. Lett., {\bf 50}, 594 (2000).

\bibitem{clade01}
S. Douady, B. Andreotti, A. Daerr and P. Clad\'e, \lq\lq~The four fronts and the two avalanches~\rq\rq, in {\em Powder and Grains 2001}, Y. Kishino ed., 443, (A.A. Balkema Publishers, Amsterdam) (2001).

\end{thebibliography}
\end{document}